\documentclass[12pt,preprint]{aastex}
\usepackage{rotating}
\usepackage{amsmath}

\newcommand{\citar}[1]{\citeauthor{#1} (\citeyear{#1})}

\newcommand{\citartwo}[2]{\citeauthor{#1} (\citeyear{#1}, \citeyear{#2})}

\newcommand{\citarNP}[1]{\citeauthor{#1} \citeyear{#1}}

\newcommand{\citartwoNP}[2]{\citeauthor{#1} \citeyear{#1}, \citeyear{#2}}
\newcommand{\citarthreeNP}[3]{\citeauthor{#1} \citeyear{#1}, \citeyear{#2}, \citeyear{#3}}

\newcommand{\threej}[6]
{\begin{pmatrix}
#1&#2&#3\\
#4&#5&#6
\end{pmatrix}}

\newcommand{\rotmat}[3]{\mathcal{D}^{#1}_{#2 #3}}

\begin{document}
\title{Dichroic Masers due to Radiation Anisotropy and the Influence of the Hanle Effect on the Circumstellar SiO Polarization}

\author{A. Asensio Ramos}
\affil{Istituto Nazionale di Astrofisica (INAF) Osservatorio Astrofisico di Arcetri, Largo Enrico Fermi 5, 50125 Florence, Italy}
\author{E. Landi Degl'Innocenti}
\affil{Dipartimento di Astronomia e Scienza dello Spazio, Largo Enrico Fermi 2, 50125 Florence, Italy}
\and
\author{J. Trujillo Bueno\altaffilmark{1}}
\affil{Instituto de Astrof\'{\i}sica de Canarias, 38205, La Laguna, Tenerife, Spain}
\altaffiltext{1}{Consejo Superior de Investigaciones Cient\'{\i}ficas, Spain}
\email{aasensio@arcetri.astro.it,landie@arcetri.astro.it,jtb@iac.es}

\begin{abstract}
The theory of the generation and transfer of polarized radiation, mainly developed for interpreting solar spectropolarimetric
observations, allows to reconsider, in a more rigorous and elegant way, a physical mechanism that has been suggested some years
ago to interpret the high degree of polarization often observed in astronomical masers. This mechanism, for which the name of
``dichroic maser'' is proposed, can operate when a low density molecular cloud is illuminated by an anisotropic source of radiation
(like for instance a nearby star). Here we investigate completely unsaturated masers and show that
selective stimulated emission processes are capable of producing
highly polarized maser radiation in a non-magnetic environment. The
polarization of the maser radiation is linear and is directed tangentially to a ring equidistant to the central star. We show that the
Hanle effect due to
the presence of a magnetic field can produce a rotation (from the tangential direction) of the polarization by
more that 45$^\circ$ for some selected combinations of the strength, inclination and azimuth of the magnetic field vector. However, these
very same conditions produce a drastic inhibition of the maser effect. The rotations of about 90$^\circ$ observed in SiO masers
in the evolved stars TX Cam by \citar{kemball_diamond97} and IRC+10011 by \citar{desmurs00} may then be explained
by a local modification of the anisotropy of the radiation field, being transformed from mainly radial to mainly tangential.
\end{abstract}


\section{Introduction}
Growing attention has been devoted in recent years to the study of non-equilibrium phenomena involving populations of magnetic
sublevels in astrophysical plasmas. Most of the work in this field has been carried out within the framework of the quantum theory
of spectral line polarization
and has been aimed at the physical understanding and
numerical modeling of the
scattering polarization phenomena observed in the radiation emitted by the
outer layers of stellar atmospheres and, more particularly, of the solar atmosphere (\citarNP{trujillo_landi97},
\citartwoNP{landi98}{landi03a}, \citarthreeNP{trujillo99}{trujillo01}{trujillo03}, \citarthreeNP{trujillo02a}{trujillo02b}{trujillo04},
\citarNP{casini02}, \citarNP{manso02}, \citarNP{manso03}, \citarNP{landi04}).

Due mainly to limb darkening and to geometrical effects, the radiation field in the outer layers of stellar atmospheres is anisotropic
and therefore capable of
introducing differences in the populations of the magnetic sublevels (either degenerate or split by a magnetic field). These populations
imbalances, in some cases accompanied by more complicated phenomena such as coherences --or
quantum interferences-- between different magnetic
sublevels, are responsible for the appearance of polarization in the radiation emitted by the atoms (resonance polarization). The
phenomenon of population imbalances between magnetic sublevels is well known in laboratory spectroscopy where it is often referred
to as ``atomic polarization'' (\citarNP{happer72}).
A low density plasma irradiated by a directional source of radiation --like for instance a nearby star-- is the most appropriate
physical environment where atomic polarization can play a fundamental role. This is because the atoms and/or molecules
are irradiated by a highly anisotropic radiation field, whereas collisions with nearby particles are very sparse
and are not capable of destroying the atomic polarization induced by the radiation.

In order to get a qualitative idea of the phenomena that we are going to address in this paper, consider two atomic (or molecular) levels
that are very close in energy like, for instance, two successive rotational levels of a particular vibrational state of a molecule,
and suppose that the average population of the upper level (defined as the overall population divided by the statistical weight) is
slightly smaller than the one of the lower level. In this situation, no masing action is possible according to the conventional
view. But if a certain amount of atomic polarization is present either in the lower or in the upper level (or in both), it may
well happen that \emph{some magnetic sublevels of the upper level turn out to have a population larger than some magnetic sublevels
of the lower level}, so that a particular kind of population inversion will exist between the two levels. This phenomenon will be
referred to in the following as ``selective population inversion''. In this situation, masing action may become possible, though
only between specific magnetic sublevels, and the result is that a radiation of a particular polarization character is amplified
by stimulated emission, whereas radiation having a different polarization character is absorbed. A collection of atoms or molecules
with selective population inversion thus behaves as a dichroic medium of a particular nature\footnote{A medium is said to be dichroic
when its absorption coefficient depends on the polarization of the incident radiation. In our case, the concept of dichroism can be
generalized to allow also for stimulated emission effects.}.
For this reason, we will refer to this
kind of masing action as ``dichroic masing''. Phenomenological versions of this mechanism have indeed been invoked in the past to explain
the large degree of
polarization observed in astronomical SiO masers (\citarthreeNP{western_watson83a}{western_watson83b}{western_watson84}). Detailed
calculations have been performed for the simplest transitions in saturated masers ($J = 1 \to 0$, $J = 2 \to 1$) and have led
to interesting results.

Recent VLBA observations of SiO masers in the circumstellar environment of late-type stars
have shown that the linear polarization of the SiO maser transitions
is mainly tangential, i.e. in the direction parallel to a ring at a given distance to the central star (\citarNP{kemball_diamond97}).
These observations,
complemented by further observations of circular polarization, were interpreted, using a formula by \citar{elitzur96}, as due to the
presence of a strong magnetic field of $\sim 10$ G.
\citar{desmurs00} noted that this magnetic field appears to be
uncomfortably strong. They also noted that \citar{kemball_diamond97} found difficulties for defining a topological distribution
of magnetic fields which was able to explain the tangential polarization. By means of a phenomenological approach,
\citar{desmurs00} suggest that the radiative pumping
mechanism can easily explain the tangential polarization observed in the SiO maser spots as a result of the anisotropic radiation
field coming from the star which is illuminating the spot, even in the absence of magnetic fields. This mechanism, based on the assumption
of a difference in the pumping rates for the different magnetic sublevels, was first proposed, though in a phenomenological way,
by \citar{bujarrabal_nguyen81} and later investigated in detail by \citartwo{western_watson83a}{western_watson83b}. Even more interesting
is the fact that the observations by \citar{kemball_diamond97} show some SiO maser spots in which the polarization direction is almost radial instead of being tangential.

The quantum theory of polarization in spectral lines (see the monograph by \citarNP{landi04})
is capable of dealing with this topic in a more rigorous and elegant way. Here, we apply it to the investigation of the polarization
properties of the SiO maser lines in the unsaturated regime and of the mechanisms which may produce a rotation of the direction of
polarization, i.e., the Hanle effect due to
the presence of a magnetic field and/or a local variation of the anisotropy properties of the radiation field. Some of the results
presented in this paper, in particular, those contained in Sec. \ref{sec_maser_condition} have been previously obtained by
\citar{litvak75} but we consider our formulation to be more rigorous.
Moreover, it contains in a self-consistent manner all the physical mechanisms which may introduce population inequalities and coherences
among the magnetic sublevels. These physical mechanisms are known (e.g., \citartwoNP{western_watson83a}{western_watson83b}) but they
have always been treated in a phenomenological manner.

\section{Basic equations}

\subsection{Polarized Radiative Transfer}
Consider the propagation of a beam of polarized radiation through a medium which is optically active in the sense that its optical
properties are such that they can generate and modify the polarization of the radiation. The propagation is
described by the following radiative transfer equation (\citarNP{landi83}):
\begin{equation}
\frac{d}{ds} \mathbf{I} = \mathbf{\epsilon} - \mathbf{K} \mathbf{I},
\label{eq_radiative_transfer}
\end{equation}
where $\mathbf{I}=(I,Q,U,V)^\dag$ is the Stokes vector
at the frequency and propagation
direction under consideration (with $\dag$ indicating the transpose of the vector),
$\mathbf{K}$ is the 4$\times$4 propagation matrix,
$\mathbf{\epsilon}=(\epsilon_I,\epsilon_Q,\epsilon_U,\epsilon_V)^\dag$ is
the emission vector in the four Stokes parameters and $s$ is the geometrical distance along the ray. The $\mathbf{K}$ matrix
contains contributions from both absorption and
stimulated emission processes, which can be labeled as $\mathbf{K}^A$ and $\mathbf{K}^S$, respectively. The explicit form of the
matrix is the following:
\begin{equation}
\mathbf{K} = \mathbf{K}^A - \mathbf{K}^S =
\left(%
\begin{array}{cccc}
  \eta_I & \eta_Q & \eta_U & \eta_V \\
  \eta_Q & \eta_I & \rho_V & -\rho_U \\
  \eta_U & -\rho_V & \eta_I & \rho_Q \\
  \eta_V & \rho_U & -\rho_Q & \eta_I \\
\end{array}%
\right),
\label{eq_propagation_matrix}
\end{equation}
where $\eta_I=\eta_I^{(A)}-\eta_I^{(S)}$, $\eta_Q=\eta_Q^{(A)}-\eta_Q^{(S)}$, $\cdots$.
The propagation matrix
$\bf K$ can also be written as

\begin{eqnarray}
\mathbf{K} \, = \,
\left( \begin{array}{cccc}
{\eta_I}&{0}&{0}&{0} \\
{0}&{\eta_I}&{0}&{0} \\
{0}&{0}&{\eta_I}&{0} \\
{0}&{0}&{0}&{\eta_I}
\end{array} \right)\,+\,
\left( \begin{array}{cccc}
{0}&{\eta_Q}&{\eta_U}&{\eta_V} \\
{\eta_Q}&{0}&{0}&{0} \\
{\eta_U}&{0}&{0}&{0} \\
{\eta_V}&{0}&{0}&{0}
\end{array} \right)\,+\,
\left( \begin{array}{cccc}
{0}&{0}&{0}&{0} \\
{0}&{0}&{\rho_V}&{-\rho_U} \\
{0}&{-\rho_V}&{0}&{\rho_Q} \\
{0}&{\rho_U}&{-\rho_Q}&{0}
\end{array} \right),
\end{eqnarray}
which helps to clarify that it has six contributions: three due to transitions
from the lower level ($l$) to the upper level ($u$), and three due
to the stimulated emission transitions from the upper level ($u$)
to the lower level ($l$). Concerning
the contributions of $l{\rightarrow}u$ transitions, we have
{\em absorption} (the first matrix, ${\bf K}_1^{A}$,
which is responsible for the
attenuation of the radiation beam irrespective of its polarization state),
{\em dichroism} (the second matrix, ${\bf K}_2^{A}$, which accounts for a selective
absorption of the different polarization states),
and {\em anomalous dispersion} (the third matrix, ${\bf K}_3^{A}$, which describes the dephasing of the different
polarization states as the radiation beam propagates through the medium).
Concerning the contributions of $u{\rightarrow}l$ stimulated transitions we have that
${\bf K}_1^{S}$ would be the amplification matrix (responsible
for the amplification of the radiation beam irrespective of its polarization state),
${\bf K}_2^{S}$ would be the dichroism amplification matrix (responsible of a selective
stimulated emission of different polarization states), and ${\bf K}_3^{S}$
would be the anomalous dispersion amplification matrix.

\subsection{Dichroic Maser Condition}
\label{sec_maser_condition}
It can be shown that the eigenvalues of the propagation matrix are given by (see \citarNP{landi_landi85}):
\begin{equation}
\lambda_1=\eta_I - \Lambda_+, \qquad \lambda_{2,3}=\eta_I \pm \mathrm{i} \Lambda_{-}, \qquad \lambda_4=\eta_I+\Lambda_+,
\label{eq_eigenvalues}
\end{equation}
where $\mathrm{i}$ is the imaginary unit and
\begin{equation}
\Lambda_{\pm} = \sqrt{\frac{1}{2} \left[ \sqrt{(\eta^2-\rho^2)^2 + 4 (\vec \eta \cdot \vec \rho})^2 \pm (\eta^2-\rho^2) \right]},
\end{equation}
and where we have introduced the formal vectors $\vec \eta=(\eta_Q,\eta_U,\eta_V)$ and $\vec \rho=(\rho_Q,\rho_U,\rho_V)$.

A dichroic maser can occur when at least one of the eigenvalues is negative (Landi Degl'Innocenti 2003$b$).
Since the $\Lambda_{\pm}$ quantities are always non-negative,
the smallest eigenvalue is $\eta_I-\Lambda_+$. This eigenvalue can become negative in the less restrictive conditions. When this
condition is fulfilled, the particular mode of polarization associated with
this eigenvalue (namely, an eigenvector of the $\mathbf{K}$ matrix) is exponentially amplified in the medium.
It can be verified that this mode of polarization will be 100\% polarized, when amplified, if the physical conditions in the medium
remain constant and if the maser remains in the unsaturated regime.
The transition to
saturation is usually very fast, so that the optical depth is not so large as to permit one of the modes to completely dominate the
radiation (\citarNP{western_watson84}).
The other eigenvalues can become negative under more restrictive conditions. The mode of polarization associated
with $\lambda_2$ and $\lambda_3$ becomes amplified when $\eta_I<0$ while that associated with $\lambda_4$ becomes amplified when
$\eta_I+\Lambda_+<0$. In any case, the mode associated with the minimum eigenvalue, $\eta_I-\Lambda_+$,
will be preferentially amplified.

In the usual case where polarization phenomena are neglected, the matrix $\mathbf{K}$ reduces to the identity matrix multiplied
by the standard absorption coefficient (including the contribution from stimulated emission). Therefore, the quantities $\Lambda_{\pm}$
are zero and all the eigenvalues of the propagation matrix reduce to a single one, equal to $\eta_I$. In this case, the population
inversion requirement is $\eta_I<0$ so that the amplification is independent of the polarization mode of the radiation.

Consider an atomic system (atom or molecule) in an anisotropic environment having cylindrical symmetry around a given direction that
we choose as the quantization axis of total angular momentum
(the $z$-axis of our reference system). Consider the case where there is no magnetic field present in
the medium. The populations of the single magnetic sublevels can be deduced by solving the statistical equilibrium equations for the
multilevel atom case discussed in sections \emph{7.1} and \emph{7.2} of
\citar{landi04}. Such equations contain radiative rates and collisional rates due to collisions with the surrounding
particles, whose velocity distribution can be considered, in a broad range of physical conditions, to be isotropic. It can be shown
that, in such a physical environment, the atomic
system can be described, using the formalism of the irreducible statistical tensors $\rho^K_Q(\alpha,J)$ (see, e.g., \citarNP{landi04}),
by the only elements
with $K$ even and $Q=0$. Note that the statistical tensors can also be referred as the multipole components of the atomic density matrix.
In this formalism, $\rho^0_0(\alpha,J)$ is proportional to the total population of the level with total angular
momentum $J$, while $\rho^2_0(\alpha,J)$ is
the so-called alignment coefficient. $\alpha$ is a collection of inner quantum numbers which include, among others, the vibrational
quantum number. From now on, we will drop all the inner quantum numbers except for the vibrational one, $v$, when needed. For instance, for
a level with $J=1$, one has:
\begin{align}
\rho^0_0(J=1) &= (n_{-1}+n_0+n_1)/\sqrt{3} \\
\rho^2_0(J=1) &= (n_{-1}-2n_0+n_1)/\sqrt{6},
\end{align}
where $n_M$ is the population of each magnetic sublevel $M=-1,0,1$.

Consider an electric dipole transition between two levels, the lower level $(v_l,J_l)$ and the upper level $(v_u,J_u)$. The
elements of the propagation matrix $\mathbf{K}$ can be expressed in terms of the statistical tensors $\rho^K_0(v_l,J_l)$ which
account for the absorption processes and $\rho^K_0(v_u,J_u)$ which account for the stimulated emission processes. The general expressions
of the elements of the propagation matrix can be found in \citar{landi04},
while the explicit expressions for the particular case of a line transition without
overlapping with other lines can be found in \citar{trujillo03}. Consider the scattering geometry shown in
Fig. \ref{fig_geometry}. For a given
direction $\mathbf{\Omega}$, passing through the atom or molecule and
forming angles $\theta$ (polar angle, usually parameterized as $\mu=\cos \theta$) and $\chi$ (azimuth) with the quantization
axis $z$, and defining the positive $Q$-direction in the plane $(z,\mathbf{\Omega})$ (vector $\mathbf{e}_1$ in the figure),
all the elements of the propagation matrix are zero, except for $\eta_I$, $\eta_Q$ and $\rho_Q$.
This is valid for the
zero magnetic field case that we are considering in this section.
For this particular case, $\Lambda_{+}$ and $\Lambda_{-}$ turn out to be simply given by $|\eta_Q|$ and $|\rho_Q|$, respectively.
Therefore, the selective masing condition is
\begin{align}
\lambda_1=\eta_I-|\eta_Q|=\frac{h \nu_0}{4 \pi} \mathcal{N} (2J_l+1) B(J_l \to J_u) \Bigg[ a + b(3\cos^2\theta - 1) - 3|b|\sin^2\theta
\Bigg] \phi(\nu_0-\nu) < 0,
\end{align}
where $\mathcal{N}$ is the number density of molecules, $B(J_l \to J_u)$ is the Einstein coefficient for absorption of the transition,
and $\phi(\nu_0-\nu)$ is the line profile centered at the transition frequency $\nu_0$ and normalized to unity in frequency. The
quantities $a$ and $b$ are given by:
\begin{gather}
a = \frac{\rho^0_0(J_l)}{\sqrt{2J_l+1}} - \frac{\rho^0_0(J_u)}{\sqrt{2J_u+1}} \nonumber \\
b = \frac{1}{2 \sqrt{2}} \left[ w^{(2)}_{J_l J_u} \frac{\rho^2_0(J_l)}{\sqrt{2J_l+1}} -
w^{(2)}_{J_u J_l} \frac{\rho^2_0(J_u)}{\sqrt{2J_u+1}} \right],
\end{gather}
with the symbols $w^{(2)}_{J_l J_u}$ and $w^{(2)}_{J_u J_l}$ defined in \citar{landi84}. Depending on the sign of $b$, the masing condition
can be expressed as:
\begin{equation}
\begin{cases}
a + 2b < 0 & \text{if $b<0$} \\
a + [3\cos (2\theta)-1]b < 0 & \text{if $b>0$}.
\label{eq_maser_conditions}
\end{cases}
\end{equation}

It can be verified that the direction of polarization of the emergent maser radiation is tangential
to a ring at a given distance to the
central star if $b<0$, while it is perpendicular if $b>0$. Of course, when $b=0$, there is no atomic polarization, and the maser conditions
transform into the usual condition $a<0$. Finally, it is interesting to note that, even if $a>0$, we can have a maser effect due to the
presence of atomic polarization in the rotational levels, as already discussed in the Introduction.

\subsection{SiO molecule}
The results derived in the previous section can be applied to a simple SiO molecular model. This model represents the two lowest
vibrational levels of SiO ($v=0$ and $v=1$) in the fundamental electronic state $X^1\Sigma^+$. The number of rotational levels included
in the model does not need to be specified. For simplicity, we assume that the population of the energy levels of the
fundamental vibrational level
are in thermal equilibrium at a collisional temperature $T_c$ and that the population of the rotational levels of the $v=1$
vibrational level is due to the pumping of the stellar infrared radiation at 8 $\mu$m. The effect of collisions on the $v=1$ levels
is neglected.

Since the rotational levels of the lowest vibrational level are thermalized, the statistical tensors which describe
the atomic polarization of these levels can be written as:
\begin{equation}
\rho^K_Q(v=0,J_l) = \delta_{K0} \delta_{Q0} \rho^0_0(v=0,J_l),
\end{equation}
which states that all multipoles vanish except the one with $K=0$ and $Q=0$, which represents, apart from a factor, the overall population
of the rotational level. They can be rewritten by using the Boltzmann law as
\begin{equation}
\rho^0_0(v=0,J_l) = \frac{\sqrt{2J_l+1}}{Z(T_c)} \exp[-E(v=0,J_l)/T_c].
\end{equation}
$Z(T_c)$ is the partition function at the collisional temperature $T_c$ and $E(v=0,J_l)$ are the energies of the rotational levels.

The population and polarization state of the $v=1$ levels are mainly driven by the vibro-rotational transitions, while the pure
rotational transitions are merely anecdotic. Therefore, we only include in our model the vibro-rotational transitions. Since
the electronic state has no spin and no orbital angular momentum, Hund's case (b) gives a very good approximation to the
coupling of angular momenta in the SiO molecule. The Einstein coefficients for spontaneous emission can be written as:
\begin{equation}
A(v_u,J_u \to v_l,J_l) = A_0 (2J_l+1) \threej{J_u}{J_l}{1}{0}{0}{0}^2,
\end{equation}
where $A_0=6.61$ s$^{-1}$ is the band integrated Einstein coefficient, obtained from \citar{drira97}.
The statistical tensors $\rho^K_Q(v,J)$
of the rotational levels with $v=1$ can be calculated by solving the statistical equilibrium equations
(see \citarNP{landi84} for an example).
We neglect stimulated emission produced by the infrared radiation coming from the star. This approximation is valid only if the
number of photons per mode ($\bar n$) of the pumping radiation is very small.\footnote{When this is not the case, the main effect
of stimulated emission is the one of reducing atomic polarization, mimicking the effect of a reduced anisotropy factor
(see \citarNP{landi04}, Section 10.9).}
Assuming a blackbody radiation for a central star
with a temperature $T_\star=2500$ K and affected by a geometrical dilution factor $f$ due to the distance, we get:
\begin{equation}
\bar n = \frac{c^2}{2 h \nu^3} B(T_\star) f,
\end{equation}
which yields $\bar n < 0.1$ at 8 $\mu$m for distances of the order or larger than 2 stellar radii.
Under such circumstances, the statistical equilibrium equations can be written as:
\begin{align}
\frac{d}{dt} \rho^K_Q(v=1,J_u) =& \sum_{J_l} \sum_{K_l Q_l} T_A(JKQ,J_lK_lQ_l) \rho^{K_l}_{Q_l}(v=0,J_l) \nonumber \\
&- \sum_{K' Q'} R_E(JKQ,K'Q') \rho^{K'}_{Q'}(v=1,J_u) = 0,
\label{eq_SEE}
\end{align}
where $T_A(JKQ,J_lK_lQ_l)$ and $R_E(JKQ,K'Q')$ are transfer and relaxation radiative rates whose expressions can be found
in section {\em 7.2.a} of \citar{landi04}.
Inserting in Eq. (\ref{eq_SEE}) the expressions for the radiative rates, after some Racah algebra we end up with the following expression:
\begin{equation}
\rho^K_0(v=1,J_u) = \sum_{J_l} \frac{c^2}{2 h \nu^3} \sqrt{(2J_l+1)(2J_u+1)} \threej{J_u}{J_l}{1}{0}{0}{0}^2 w_{J_u J_l}^{(K)}
\rho^0_0(v=0,J_l) J^K_0.
\label{eq_rhos_nofield}
\end{equation}
The quantities $J^K_Q$ are the tensors of the radiation field at the wavelength of the infrared transition. Consider a plane-parallel
layer in which the SiO molecules are present.
We select the quantization axis along the radial direction, i.e., along the perpendicular to the slab. If the radiation field due to
the star is unpolarized and axisymmetric around the quantization axis, the radiation field
can be fully described with the following two tensors:
\begin{eqnarray}
J^0_0 &=& \frac{1}{2} \int_{-1}^{1} d\mu I(\mu) \\
J^2_0 &=& \frac{1}{4\sqrt{2}} \int_{-1}^{1} d\mu \left[ (3\mu^2-1)I(\mu) \right]
\label{eq_tensors_rad_field},
\end{eqnarray}
which are frequently parameterized in terms of the number of photons per mode, $\bar n$, and the anisotropy factor, $w$, defined by
\begin{equation}
\bar n = \frac{c^2}{2h\nu^3} J^0_0, \qquad w = \sqrt{2} \frac{J^2_0}{J^0_0}.
\end{equation}
Note that the anisotropy factor is bounded in the interval $[-1/2,1]$. Both extremes
correspond to radiation predominantly directed
perpendicularly to the symmetry axis of the radiation field and to radiation predominantly directed along the symmetry axis, respectively (e.g., the review by \citarNP{trujillo01}).

The expressions derived above allows us to investigate the conditions under which a dichroic maser appears in this simplified model
of the SiO molecule. Let us assume for simplicity that we perform the observation at $\mu=0$, which represents the geometry of a
90$^\circ$ scattering. This assumption is not so restrictive since this is precisely the geometry which maximizes the optical path of the
maser radiation. In this extreme case, the maser conditions given by Eqs. (\ref{eq_maser_conditions}) transform into $a+2b<0$ if $b<0$ and
$a-4b<0$ if $b>0$.

\begin{figure}
\plottwo{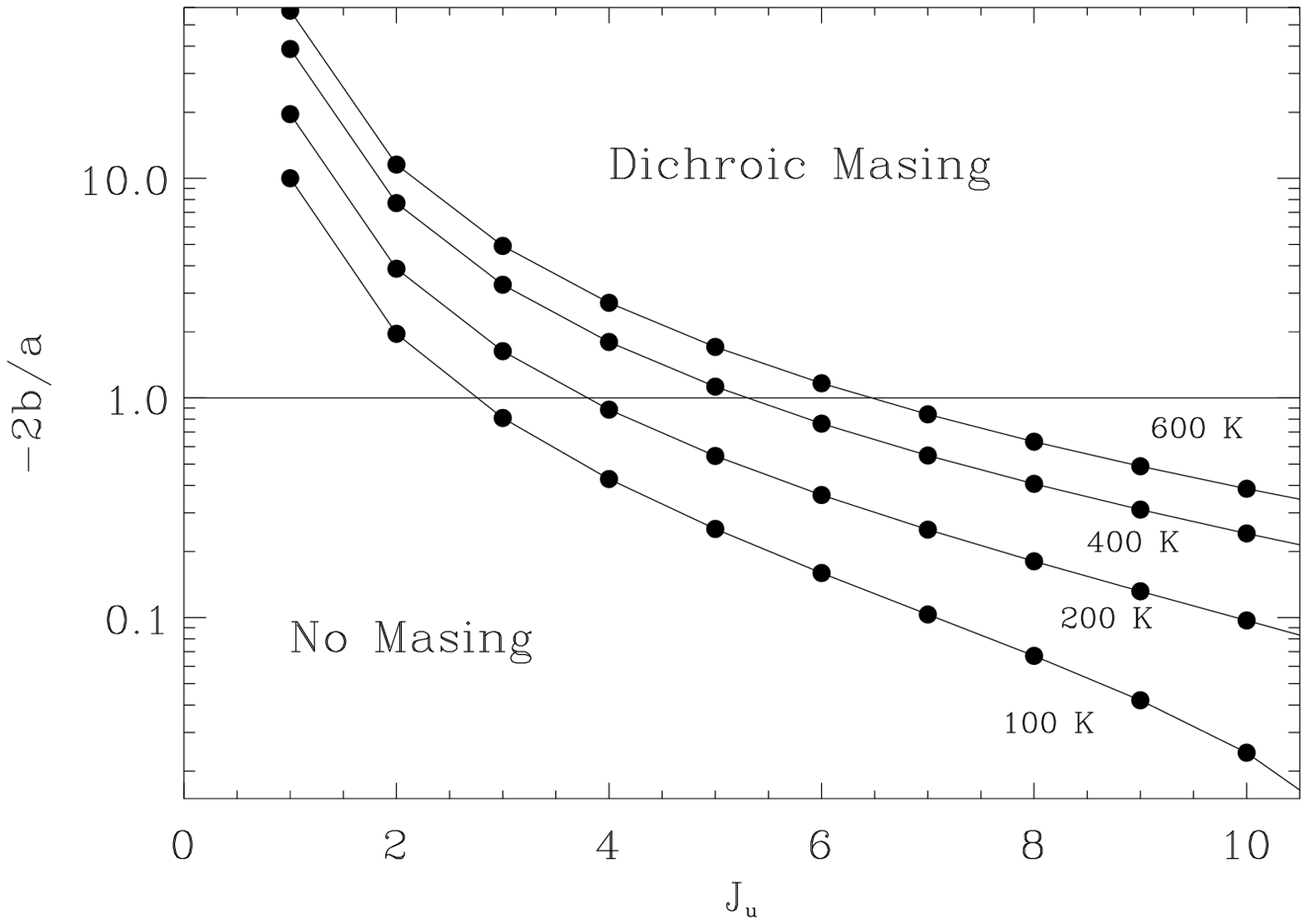}{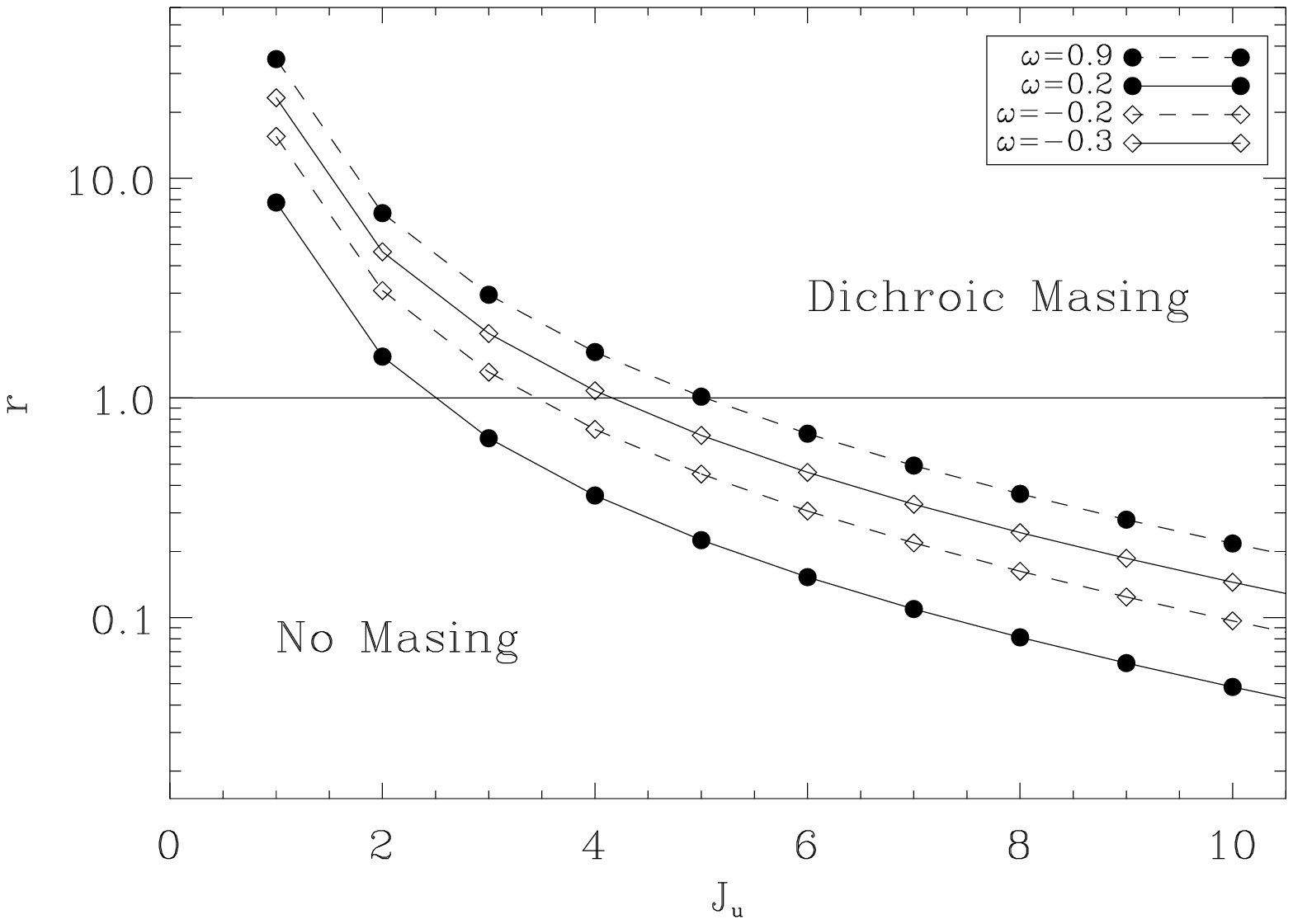}
\caption{Maser condition for the rotational transitions inside the $v=1$ vibrational level of SiO for the non-magnetic case for different
values of the collisional temperature $T_c$ (left panel) and for different
values of the anisotropy factor $w$ (right panel). The calculations of the left panel have been obtained with $w=1$. The quantity which
gives the maser condition when $b<0$ (equivalent to $w>0$ in the non-magnetic case) is $-2b/a$ while it is $4b/a$ when $b>0$ (equivalent
to $w<0$ in the non-magnetic case). We designate this quantity as $r$ in the right plot since we are plotting results with different
values of $w$. The transitions are labeled with their upper total quantum number $J_u$. The number of
inverted transitions increases when the collisional temperature is increased. It is interesting to note that the change of sign of the
anisotropy factor does not destroy the dichroic maser in the lower rotational levels.\label{fig_ratio_nofield}}
\end{figure}

Taking into account the properties of the symbols $w^{(2)}_{JJ'}$, it is easily verified that the quantity $b$ has the opposite sign of $w$
for transitions involving rotational levels with relatively
low $J$ values, while the quantity $a$ is always positive. In the left
panel of Figure \ref{fig_ratio_nofield} we show the variation of the ratio $-2b/a$ for each transition having an upper level $J_u$.
This calculation has been obtained for $w=1$ so that we have used the ratio $-2b/a$ to detect the masing transitions. The
curves have been calculated with different values of the
collisional temperature $T_c$.
Note that, the higher the temperature, the larger the
number of transitions which show selective population inversion. The dependence of the population inversion with
$J$ seems to follow the general observational results that the masers get weaker the higher the value of $J$ in the $v=1$ level
(\citarNP{jewell87}, \citarNP{cerni93}, \citarNP{bujarrabal94}).
Because the model we are investigating is completely radiative (except for
the populations of the $v=0$ vibrational level), we do not obtain a
thermalization of the energy levels to this local temperature. The right panel of Figure \ref{fig_ratio_nofield} represents the
results for a fixed temperature $T_c=400$ K and for different values of $w$ which span the allowed range. In this case, since
we have positive and negative values of $b$, we have plotted the value of the masing condition appropriate in each case, i.e.,
$r=-2b/a$ if $b<0$ and $r=4b/a$ if $b>0$. The first conclusion from this plot is that the masing conditions, although sensitive
to the actual value of $w$, do not vary very much in the whole range of variation of the anisotropy factor. Of course, when $w=0$
the radiation field is isotropic and we do not obtain a dichroic maser. The most important conclusion is that, since the polarization
direction
of the amplified radiation is dictated by the sign of $b$, {\em the variation of $w$ represents the only mechanism to switch from
maser radiation which is vibrating perpendicular to the quantization axis to radiation vibrating parallel to the quantization axis}.
The variation of any other parameter included in the model only changes the excitation state of the SiO rotational levels, but
not the polarization of the amplified radiation.

\section{The influence of the Hanle effect}
We now investigate the influence of a magnetic field on the atomic polarization of the rotational levels of the SiO
molecules and on the polarization of the emergent radiation.
This is nothing, but the so-called Hanle effect (see \citarNP{trujillo01} for a recent review).

\subsection{Statistical tensors with a magnetic field}
\begin{figure}
\plotone{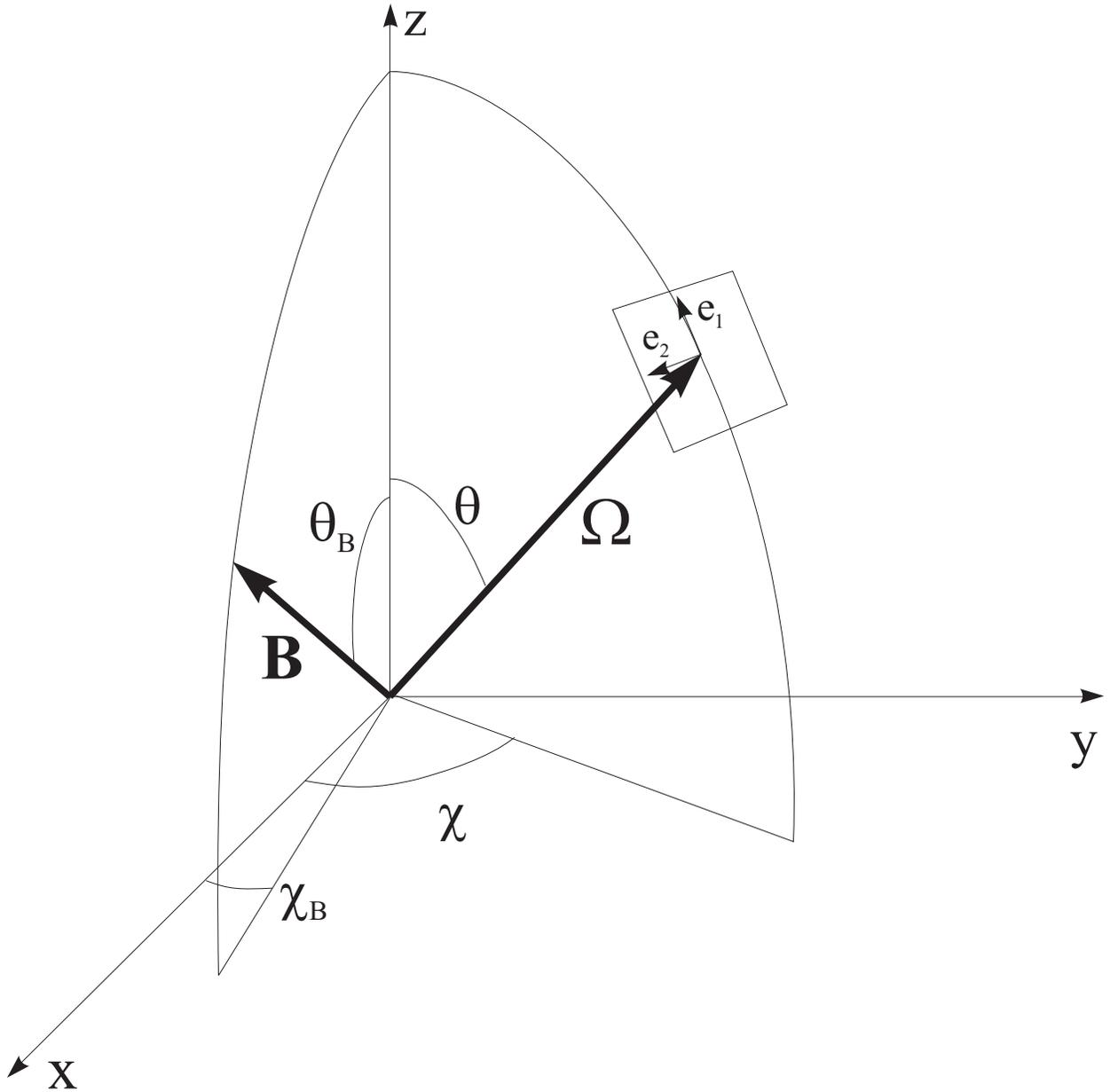}
\caption{Scattering geometry indicating the quantization axis $z$, the direction of observation $\mathbf{\Omega}$ and the magnetic
field vector $\mathbf{B}$. We have also indicated the unitary vectors $\mathbf{e}_1$ and $\mathbf{e}_2$, being the vector $\mathbf{e}_1$
the one that sets the reference direction for Stokes $Q$. \label{fig_geometry}}
\end{figure}

The presence in the maser formation region
of a magnetic field inclined with respect to the
symmetry axis of the pumping radiation field
produces a symmetry breaking. As a result,
the problem becomes much more complicated. Now
the statistical tensors with $Q \neq 0$ have to be taken into account
in order to have a correct description of the polarization state of the energy levels.
The magnetic field $\mathbf{B}$ is oriented along
a direction $\mathbf{\Omega}_B$ which, in general, does not coincide with the quantization axis (see Figure \ref{fig_geometry}).
However, in the reference system
in which the quantization axis is chosen along the direction of the magnetic field, the statistical equilibrium equations
are barely modified (see Section \emph{7.2a} of \citarNP{landi04})
and can be easily solved, obtaining:

\begin{align}
[\rho^K_Q(v=1,J_u)]_{\mathrm{mag}} = \frac{1}{1+i\Gamma Q} \sum_{J_l} \frac{c^2}{2 h \nu^3} \sqrt{(2J_l+1)(2J_u+1)}
\threej{J_u}{J_l}{1}{0}{0}{0}^2 \times \nonumber \\
(-1)^Q w_{J_u J_l}^{(K)} \rho^0_0(v=0,J_l) [J^K_Q]_\mathrm{mag},
\end{align}
where we have explicitly indicated that this is valid only in the magnetic field reference frame. Since the tensors of the radiation field
have also to be calculated in this same reference frame, we loose the cylindrical symmetry property and the tensor $[J^K_Q]_\mathrm{mag}$
contains also terms with $Q \neq 0$. The effect of the magnetic field is to produce a reduction and a dephasing of the statistical tensors with $Q \neq 0$. This reduction and dephasing in the magnetic field reference frame
depends on the function
$\Gamma$ which, in turn, depends on the magnetic field strength through the equation:
\begin{equation}
\Gamma = \frac{2\pi \nu_L g_L}{\sum_{J_l} A(J_u \to J_l)},
\end{equation}
where $\nu_L$ is the Larmor frequency which is proportional to the magnetic field strength and $g_L$ is the Land\'e factor of the rotational
level. The fundamental electronic level of SiO has neither spin nor electronic orbital angular momentum so that the Land\'e factor
includes the contribution from the coupling between the magnetic field and the rotation of the nuclei and a further contribution from
the coupling between the magnetic field and the rotation of the electronic cloud. For the level's Land\'e factor we have used the value
$g_L=-8.365 \times 10^{-5}$ (\citarNP{davis_muenter84}). In this case, being $\sum_{J_l} A(J_u \to J_l) \simeq 6.6$ s$^{-1}$, the magnetic
field strength which leads to the critical value $\Gamma=1$ is $\sim 9$ mG.
This means that in the presence of a magnetic field of $\sim 9$ mG
we should already
expect a significant modification in the emergent
linear polarization with respect to the zero field reference case\footnote{Actually, if $B_c$ is the critical field, the sensitivity range of the
Hanle effect lies between 0.1$B_c$ and 10$B_c$, approximately.
For fields stronger than about 10$B_c$ the emergent linear polarization
is sensitive only to the orientation of the magnetic field vector, but not
to its strength.}.

For our purposes, it is however more appropriate to calculate the statistical tensors in the reference system in which the quantization
axis is along the symmetry
axis of the slab. In order to do so, we have to carry out a rotation of the original reference system by the Euler angles $(0,-\theta_B,-\chi_B)$,
being $\theta_B$ and $\chi_B$ the angles which define the direction of the magnetic field with respect to the symmetry axis of the
slab (see Figure \ref{fig_geometry}). Taking into account the spherical tensor transformations under a rotation of the reference system, we
obtain that the statistical
tensors in the reference system with the quantization axis along the symmetry axis of the radiation field are given by the equation
\begin{equation}
[\rho^K_Q(v=1,J_u)]_{\mathrm{rad}} = \sum_{J_l} \frac{c^2}{2 h \nu^3} \sqrt{(2J_l+1)(2J_u+1)}
\threej{J_u}{J_l}{1}{0}{0}{0}^2 w_{J_u J_l}^{(K)} \rho^0_0(v=0,J_l) H^{(K)}_Q(\theta_B,\chi_B),
\end{equation}
where
\begin{equation}
H^{(K)}_Q(B,\theta_B,\chi_B) = \sum_{Q'} \frac{1}{1+\mathrm{i}\Gamma Q'} \sum_{Q''} [J^K_{-Q''}]_\mathrm{rad} (-1)^{Q''}
{\rotmat{K}{Q'}{Q}}^*(-\theta_B,-\chi_B) \rotmat{K}{Q'}{Q''}(-\theta_B,-\chi_B),
\end{equation}
$\mathcal{D}$ being the usual rotation matrices (see, e.g., \citarNP{edmonds60}).
We can simplify the previous equation taking into account that the
pumping radiation field is
unpolarized and has azimuthal symmetry, so that $J^K_Q=J^K_Q \delta_{Q0}$. In this case:
\begin{equation}
H^{(K)}_Q(B,\theta_B,\chi_B) = [J^K_{0}]_\mathrm{rad} \sum_{Q'} \frac{1}{1+\mathrm{i}\Gamma Q'} {\rotmat{K}{Q'}{Q}}^*(-\theta_B,-\chi_B)
\rotmat{K}{Q'}{0}(-\theta_B,-\chi_B).
\end{equation}
Furthermore, introducing the \emph{magnetic kernel} $\mathcal{M}^K_{QQ'}(\vec{B})$ defined by \citar{landi04}, the above expression can
be written as:
\begin{equation}
H^{(K)}_Q(B,\theta_B,\chi_B) = \mathcal{M}^K_{Q0}(\vec{B}) [J^K_{0}]_\mathrm{rad},
\end{equation}
so that the statistical tensors when a magnetic field is present can be expressed in terms of those relative to the case with $B=0$ through
the equation
\begin{equation}
[\rho^K_Q(v=1,J_u)]_\mathrm{rad}= \mathcal{M}^K_{Q0}(\vec{B}) [\rho^K_0(v=1,J_u)]_\mathrm{B=0}.
\end{equation}

We can now use some of the relevant properties of the magnetic kernel in order to gain some information on the behavior of the
statistical tensors when a magnetic field is included. Firstly, we can calculate the effect of a magnetic field on the
$\rho^0_0$ statistical tensors. In this case, we have $\mathcal{M}^0_{00}=1$ because
$\rotmat{0}{Q'}{0}=\delta_{Q' 0}$. This result means that the presence of a magnetic field does not change the overall population
of a rotational level $J$. Secondly, in the case of zero magnetic field we have to recover the original equations. Since the magnetic
kernel transforms into $\delta_{Q0}$ in the limit $\Gamma=0$, it can be proved that the equations are transformed immediately into the
zero-field ones (as a consequence of the orthogonality of the rotation matrices).
Finally, the dependence of $\mathcal{M}^K_{Q0}$ on the azimuth of the magnetic field ($\chi_B$) is periodic, with a period which
is proportional to $Q$. This can be proved by recalling the expression of the magnetic kernel in terms of the reduced rotation
matrices (\citarNP{landi04}):
\begin{equation}
\mathcal{M}^K_{Q0}(B,\theta_B,\chi_B) = e^{-\mathrm{i}\chi_B Q} \sum_{Q''} \frac{1}{1+\mathrm{i}\Gamma Q''} {d^{K}_{QQ''}}(\theta_B)
d^{K}_{Q''0}(-\theta_B).
\end{equation}
Therefore, the tensors $\rho^K_0$ are not modified when changing the azimuth of the magnetic field, and only the phase of those
with $Q \neq 0$ vary periodically with $\chi_B$.

\begin{figure}
\plotone{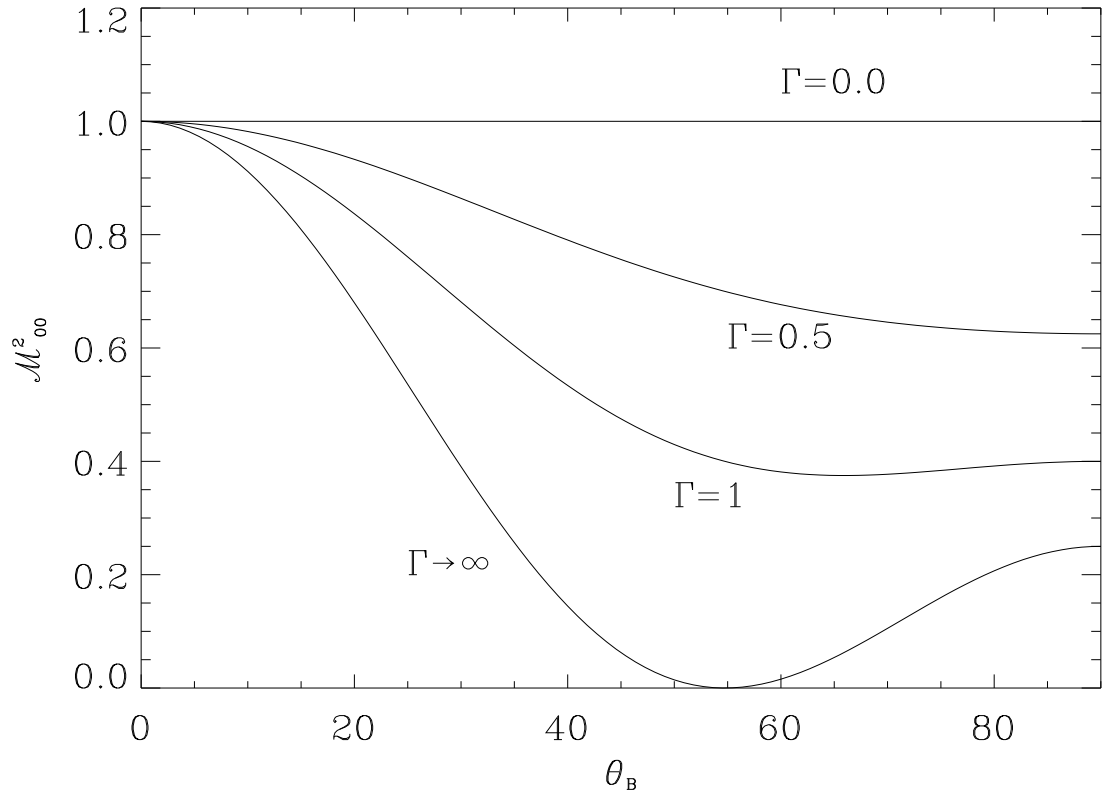}
\caption{Variation of the function $\mathcal{M}^2_{00}$ with the inclination of the magnetic field for different values of the magnetic field
intensity parameterized by the factor $\Gamma$.\label{fig_h20_angle}}
\end{figure}

From the previous equations, it is easy to see the effect of a magnetic field on the statistical tensors $\rho^2_0$, which can be
written as:
\begin{equation}
[\rho^2_0(v=1,J_u)]_\mathrm{rad}= \left[ [{d^{2}_{00}}(-\theta_B)]^2 +
\frac{2}{1+\Gamma^2} [{d^{2}_{10}}(-\theta_B)]^2 + \frac{2}{1+4\Gamma^2} [{d^{2}_{20}}(-\theta_B)]^2
\right] [\rho^K_0(v=1,J_u)]_\mathrm{B=0}.
\end{equation}
Note from the previous expression that the factor between brackets, $\mathcal{M}^2_{00}$, is always non-negative for any value of the
magnetic field vector, so that, there is no sign change in $\rho^2_0$ unless $J^2_0$ changes its sign. In Fig. \ref{fig_h20_angle}
we plot the variation of
$\mathcal{M}^2_{00}$  with the inclination of the magnetic field vector. Note that, when the magnetic field is zero or very
small (small $\Gamma$)
or when the magnetic field is along the symmetry axis of the radiation field ($\theta_B=0$), the $\rho^2_0$ tensor is not
modified. However, when the field increases until reaching the critical value $\Gamma=1$, the quantity $\mathcal{M}^2_{00}$
decreases. When the field is very large, we get the asymptotic curve labeled with $\Gamma=\infty$ in the plot, which is nothing but
$[{d^{2}_{00}}(-\theta_B)]^2$. Note that $\rho^2_0$ goes
to zero for a critical angle $\theta_\mathrm{crit}=54.73^{\circ}$, which is called Van Vleck's angle ($\cos^2 \theta_\mathrm{crit}=1/3$). On the other hand, its
asymptotical value becomes $1/4$ when $\theta_B=\pi/2$, i.e., when the field is perpendicular to the symmetry axis of the radiation field.

\subsection{Maser condition}
When a magnetic field is present, we have to take into account that $\eta_U$ may be non-zero so that the eigenvalues are modified with
respect to the non-magnetic case. Indeed, in this case, the quantity $\Lambda_+$ turns out to be
given by $\sqrt{\eta_Q^2+\eta_U^2}$. Since $\Lambda_+$ is always non-negative, even in the presence of a magnetic field, $\lambda_1$
(see Eq. (\ref{eq_eigenvalues})) is the smallest of the eigenvalues and the corresponding mode will be amplified the most.
It is not easy to find a simple analytical formula for the eigenvalues since the
expressions for $\eta_I$, $\eta_Q$ and $\eta_U$ have now contributions from $\rho^2_Q$, with $Q=0,1,2$. However, if we take into account
that the statistical tensors $\rho^K_Q$ can be written in terms of the statistical tensors for zero magnetic field $[\rho^K_0]_\mathrm{B=0}$
times the $\mathcal{M}^K_{Q0}$ factors, it is possible to write expressions for the masing condition which are very similar to those
obtained for the $B=0$ case, namely:
\begin{equation}
\begin{cases}
a + b \left( f_I+\sqrt{f_Q^2+f_U^2} \right) < 0 & \text{if $b<0$} \\
a + b \left( f_I-\sqrt{f_Q^2+f_U^2} \right) < 0 & \text{if $b>0$},
\label{eq_maser_conditions_magnetic_field}
\end{cases}
\end{equation}
where $b$ is calculated assuming no magnetic field ($a$ is immune to the magnetic field in this simplified model). The quantities
$f_I$, $f_Q$ and $f_U$
are combinations of the magnetic kernels $\mathcal{M}^2_{Q0}$ which depend on the orientation and strength
of the magnetic field ($B$, $\theta_B$ and $\chi_B$) and on the observation direction ($\theta$ and $\chi$). Their explicit
expressions can be found in App. \ref{sec_app_fQ_fU}. It is important to note that in the magnetic
case the azimuth $\chi$ of the observing direction with respect to the $z$ axis becomes relevant and that,
contrary to what happens in the zero field case, the quantity which gives the polarization state of the emerging radiation is the product
$b f_Q$, instead of $b$.

\begin{figure}
\plottwo{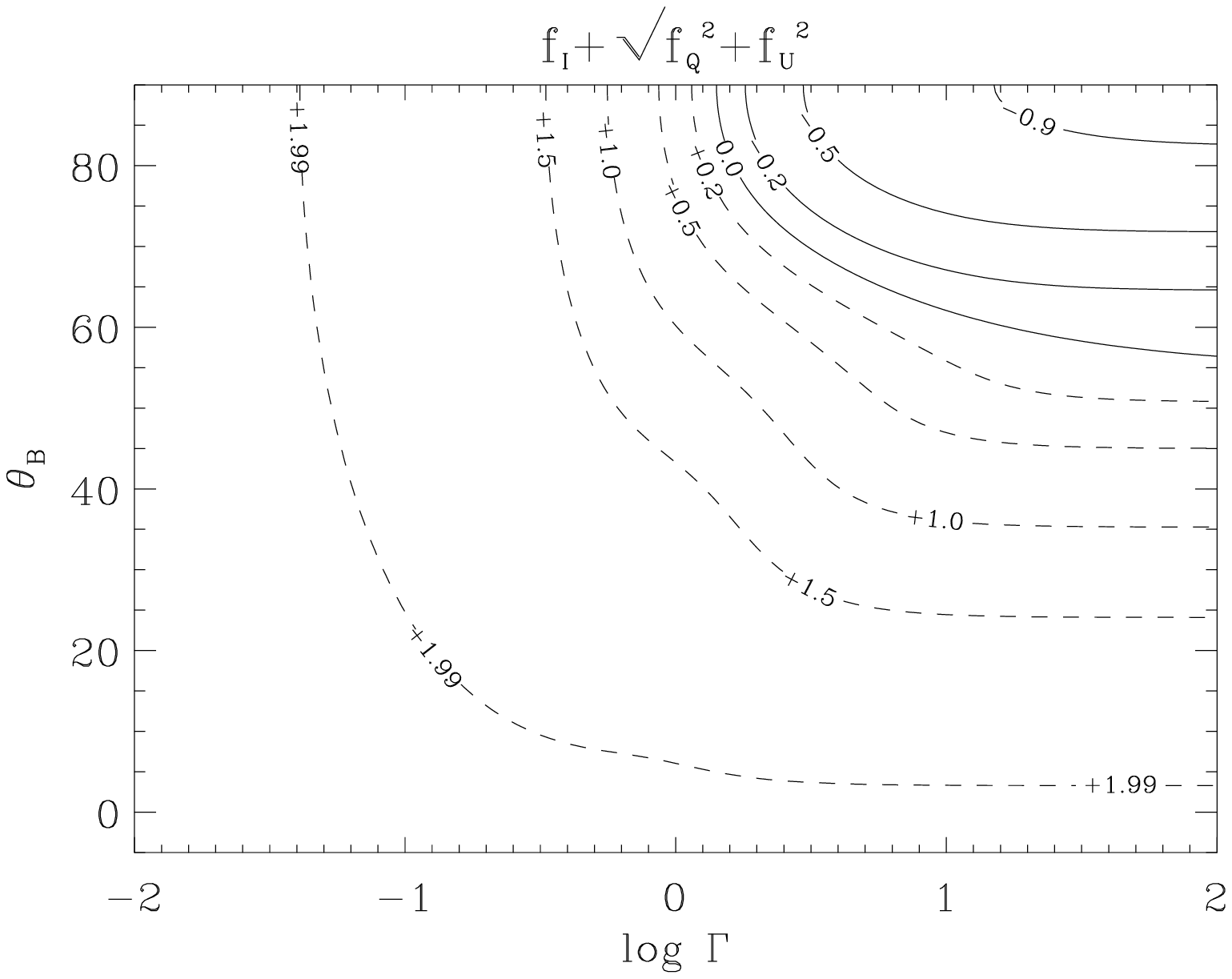}{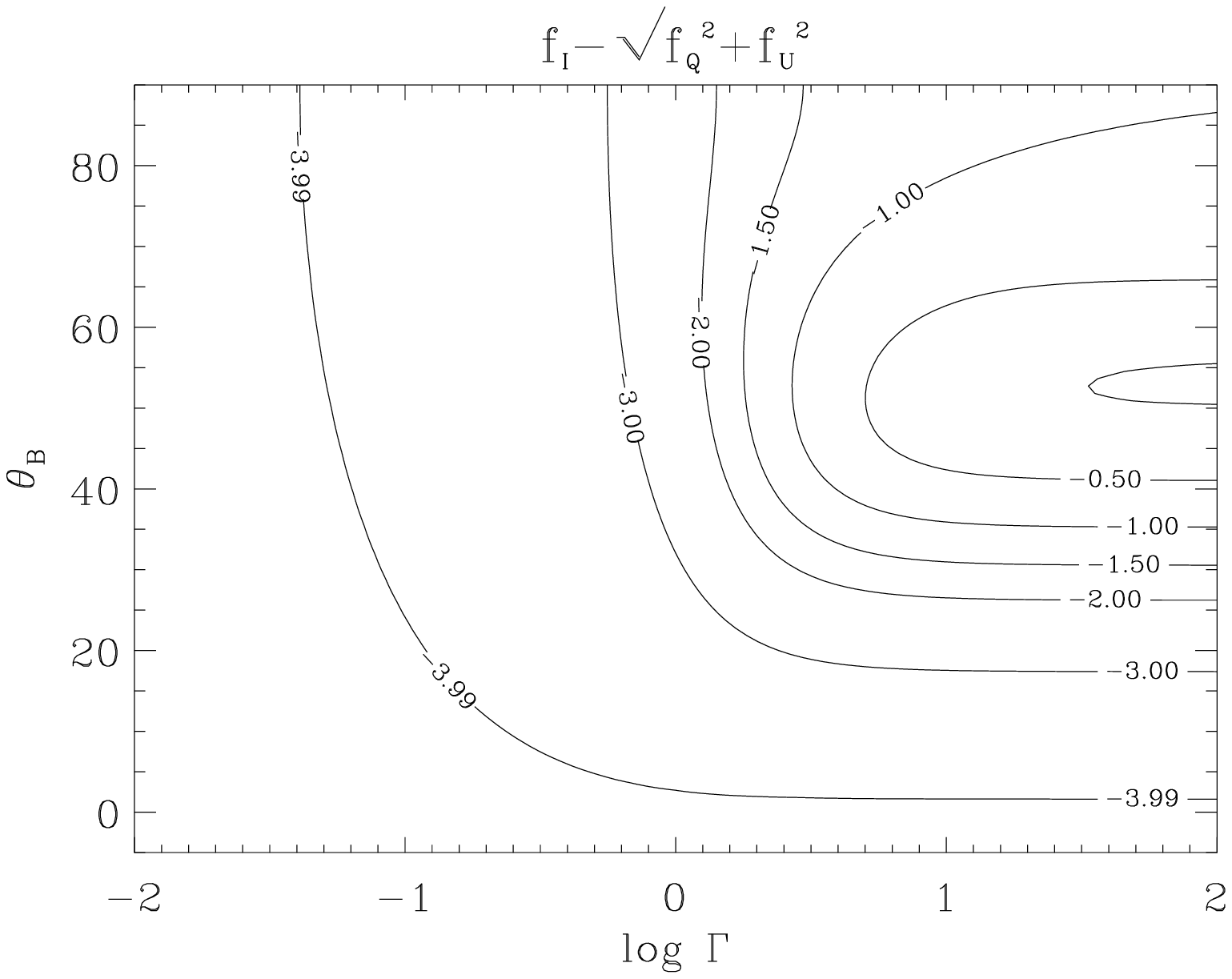}
\caption{Dependence of the factors $f_{\pm} = f_I \pm \sqrt{f_Q^2+f_U^2}$ with the strength and inclination of the magnetic field.
When the magnetic field is very weak, we recover the original values given by Eqs. (\ref{eq_maser_conditions}), namely,
$f_+ \to 2$ and $f_- \to -4$ (using $\theta=\pi/2$). When the field is
increased, the maser condition becomes more and more restrictive, until destroying any possibility of masing when the field is very
large and inclined (because $a>0$). For clarity, the isosurfaces with negative labels are plotted in solid lines, while those positive are
plotted in dashed lines.\label{fig_maser_condition_magnetic}}
\end{figure}

Assuming an observation at $\mu=0$ and $\chi=0$, we have plotted in Figure \ref{fig_maser_condition_magnetic} the variation of
the factors $f_{\pm} = f_I \pm \sqrt{f_Q^2+f_U^2}$ for different values of
the strength and inclination of the magnetic field. When the magnetic field is very weak, we recover the already discussed maser conditions,
that is, those given by Eqs. (\ref{eq_maser_conditions}) with $\theta=\pi/2$. This result is, obviously, independent of
the inclination of the magnetic field. When the field strength is increased until reaching the critical value $\Gamma=1$ or larger,
the maser conditions become more restrictive than in the zero-field case. This is true irrespectively of the sign of $b$ because both
$f_+$ and $f_-$ become smaller. The magnetic field produces an inhibition of the maser effect, leading to less inverted levels than in
the zero-field case. Focusing on the case of the SiO maser lines with $b<0$ ($w>0$), a more restrictive limiting case appears for very
inclined and strong fields for which the factor $f_+$ becomes negative. In this case, the maser mechanism has been completely inhibited
and no rotational line can be inverted. If we consider now the results with $b>0$ ($w<0$) we see that $f_-$ is always negative, except when
the inclination of the field is around the Van Vleck angle. At this angle and for fields that are strong enough, an inhibition of the
maser effect
can occur. In the rest of cases, the maser is always active.
Another interesting property is that, given an inclination of the magnetic field, a saturation
effect in the maser condition is found when the strength of the field is increased. For example, when the magnetic field has an inclination
of $\theta_B=\pi/4$, the factor $f^+$ saturates at $1/2$, one fourth of the value at $B=0$.
In the limiting case of $\Gamma \to \infty$, the expressions for $f_I$, $f_Q$ and $f_U$ simplify considerably because the magnetic
kernel does not depend on the strength of the field (in fact, $f_U=0$), so that $f_+$ and $f_-$ only depend on the inclination and azimuth
of the magnetic field and on the angles defining the line-of-sight. For $\mu=0$, $\chi=0$ and $\chi_B=0$, we obtain:
\begin{align}
f_+ = -1 + \frac{9}{2} \left[ \cos^2 \theta_B - \cos^4 \theta_B \right] +
\frac{3}{2} \sqrt{\cos^4 \theta_B \left( 3 \cos^2 \theta_B -1 \right)} \nonumber \\
f_- = -1 + \frac{9}{2} \left[ \cos^2 \theta_B - \cos^4 \theta_B \right] -
\frac{3}{2} \sqrt{\cos^4 \theta_B \left( 3 \cos^2 \theta_B -1 \right)}
\end{align}
The behavior is similar to that shown in Figure \ref{fig_maser_condition_magnetic}
for $\log \Gamma = 2$. The quantity $f_+$ is positive for field inclinations below the Van Vleck angle, negative for inclinations above this
critical angle and zero at the exact Van Vleck angle. Concerning $f_-$, it is always negative irrespective of the inclination of the
field. Only for the Van Vleck angle we find $f_-=0$.

\begin{figure}
\plottwo{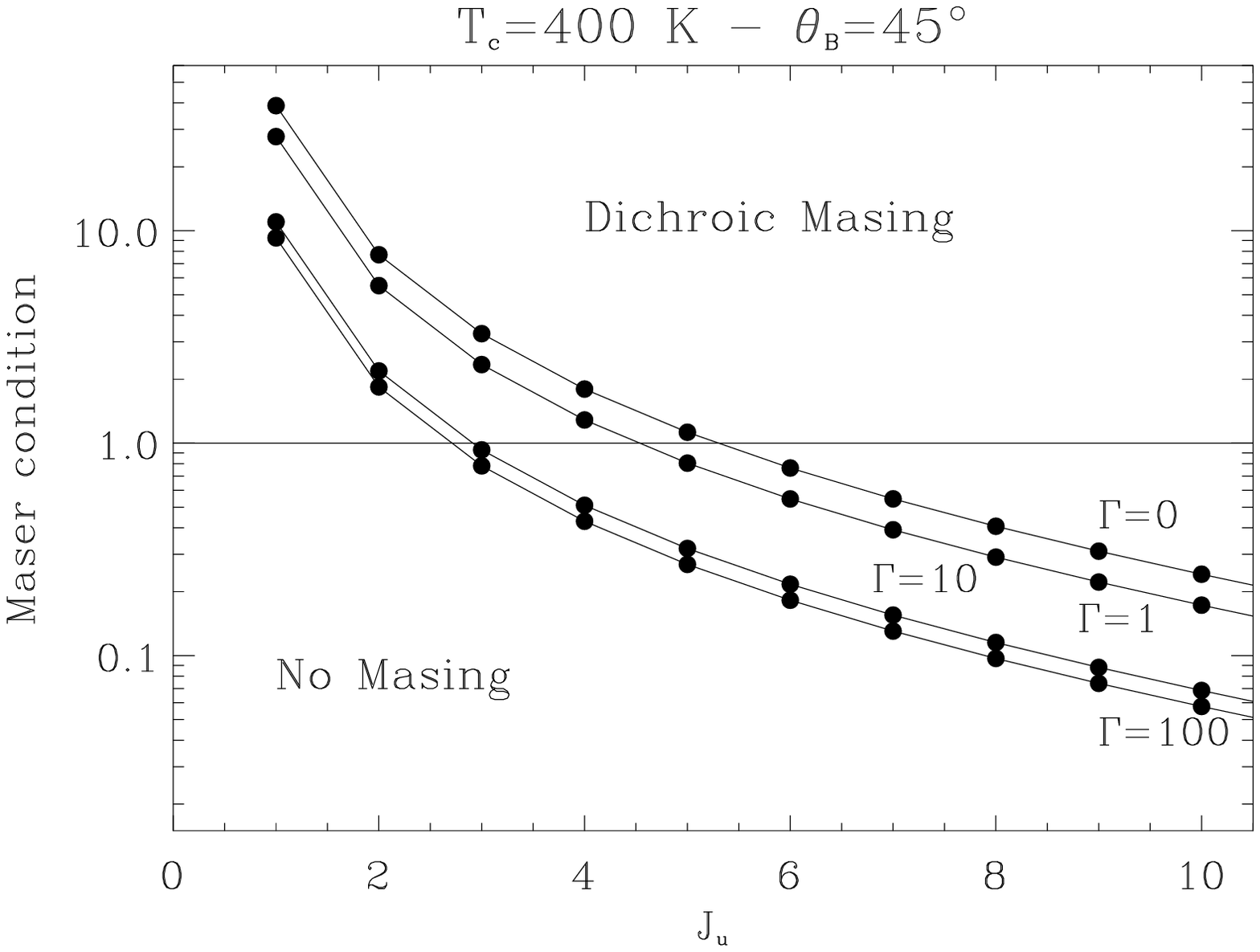}{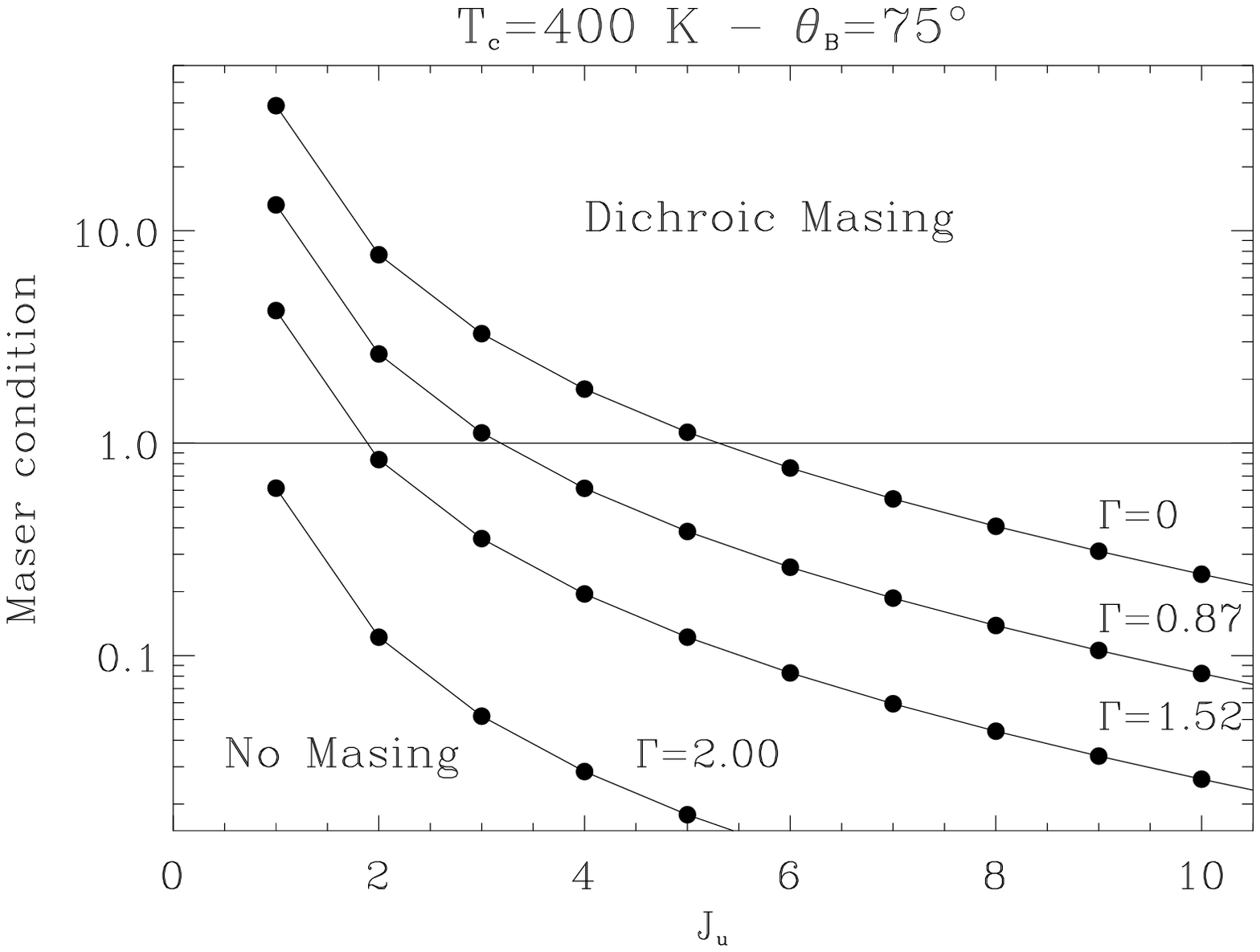}
\caption{Maser condition for two values of the inclination of the magnetic field and different values of the strength of the field.
The left panel shows that when the inclination is not very large ($\theta_B=45^\circ$, left panel) dichroic masers are possible in the lower
rotational levels even when the magnetic field strength is very large. However, when the inclination is large ($\theta_B=75^\circ$,
right panel), the dichroic maser can be completely destroyed by the magnetic field in all rotational levels. Both calculations have been
obtained using $w=1$.
\label{fig_maser_condition_magnetic_ratio}}
\end{figure}

In order to better represent the effect of a magnetic field on the maser condition, we show in Figure
\ref{fig_maser_condition_magnetic_ratio} similar plots to those of Figure \ref{fig_ratio_nofield}. We have selected a collisional
temperature $T_c$ of 400 K and two different inclinations of the magnetic field, representative of the two different behaviors which can be
seen in Figure \ref{fig_maser_condition_magnetic}. The left panel shows the results obtained when the inclination of the field is
45$^\circ$. The case $\Gamma=0$ is equal to that plotted in Figure \ref{fig_ratio_nofield} for $T_c=400$ K. Note that, when the field
is increased, the dichroic masers between the upper rotational levels of the $v=1$ vibrational level are inhibited and, in the limit of
very high magnetic fields, only the $J=1\to 0$ and $J=2\to 1$ transitions remain inverted. This is a direct consequence of the
functional form of the $f^+$ quantity shown in Figure \ref{fig_maser_condition_magnetic}. On the other hand,
on the right panel of Figure \ref{fig_maser_condition_magnetic_ratio} we have shown the results for an inclination of 75$^\circ$. In this
case, when the field strength is increased, the transitions which present a dichroic maser rapidly reduce and for the case $\Gamma=2$, we
arrive to a situation in which the maser effect has been completely inhibited by the magnetic field.

The quantities which now dictate the polarization state of the emerging maser radiation are $\eta_Q \propto f_Q b$ and
$\eta_U \propto f_U b$, where $b$ is calculated assuming zero magnetic field. However, it is difficult to obtain information
about the emergent linear polarization from the plots shown above, but we have to solve the transfer problem described by Eq.
(\ref{eq_radiative_transfer}). Taking into account the expression of the propagation matrix, neglecting the spontaneous emission term
($\mathbf{\epsilon}$) and supposing that the elements of the propagation matrix are constant along the line of sight,
the fractional linear polarization
is given, in the general magnetic case, by:
\begin{align}
Q/I = - \frac{\eta_Q}{\sqrt{\eta_Q^2+\eta_U^2}} \tanh \left( s \sqrt{\eta_Q^2+\eta_U^2} \right) \\
U/I = - \frac{\eta_U}{\sqrt{\eta_Q^2+\eta_U^2}} \tanh \left( s \sqrt{\eta_Q^2+\eta_U^2} \right).
\end{align}
When the ray proceeds through the masing region, both $Q/I$ and $U/I$ tend to an asymptotic value which is given by the ratio
in front of the hyperbolic-tangent function. Recalling the dependence of $\eta_Q$ and $\eta_U$ on the quantities $f_Q$ and $f_U$,
the asymptotic fractional linear polarizations can be expressed as:
\begin{align}
Q/I = - \frac{b}{|b|} \frac{f_Q}{\sqrt{f_Q^2+f_U^2}}\label{eq_Q_I_formal} \\
U/I = - \frac{b}{|b|} \frac{f_U}{\sqrt{f_Q^2+f_U^2}}\label{eq_U_I_formal}.
\end{align}
The sign of the fractional linear polarization is then determined by the sign of $f_Q$ and $f_U$ and by the sign of $b$. Note that these
expressions recover the sign of $Q/I$ when there is no magnetic field. In this case, $f_U=0$ and $f_Q=-3$ and $Q/I$ has the same sign
as $b$.

These expressions allow to determine the rotation angle of the linear polarization, defined by:
\begin{equation}
\alpha = \frac{1}{2} \arctan \left( \frac{U}{Q} \right) + \alpha_0,
\label{eq_rotation_angle}
\end{equation}
where $\alpha_0$ depends on the sign of $Q$ and $U$ (see, e.g., \citarNP{landi04}). From Eqs. (\ref{eq_Q_I_formal}) and
(\ref{eq_U_I_formal}), we get
\begin{equation}
\alpha = \frac{1}{2} \arctan \left( \frac{f_U}{f_Q} \right) + \alpha_0,
\label{eq_rotation_angle2}
\end{equation}

\begin{figure}
\plottwo{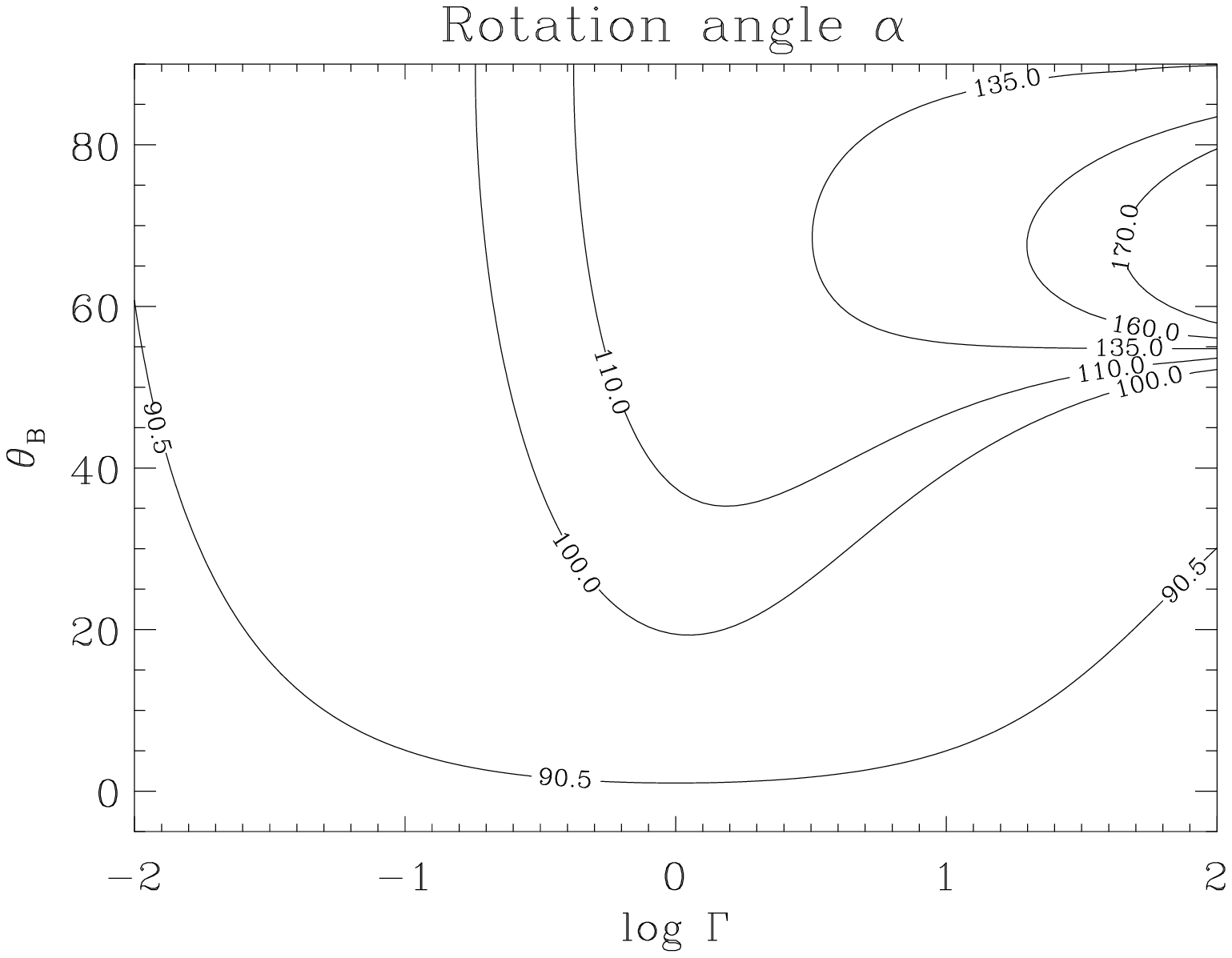}{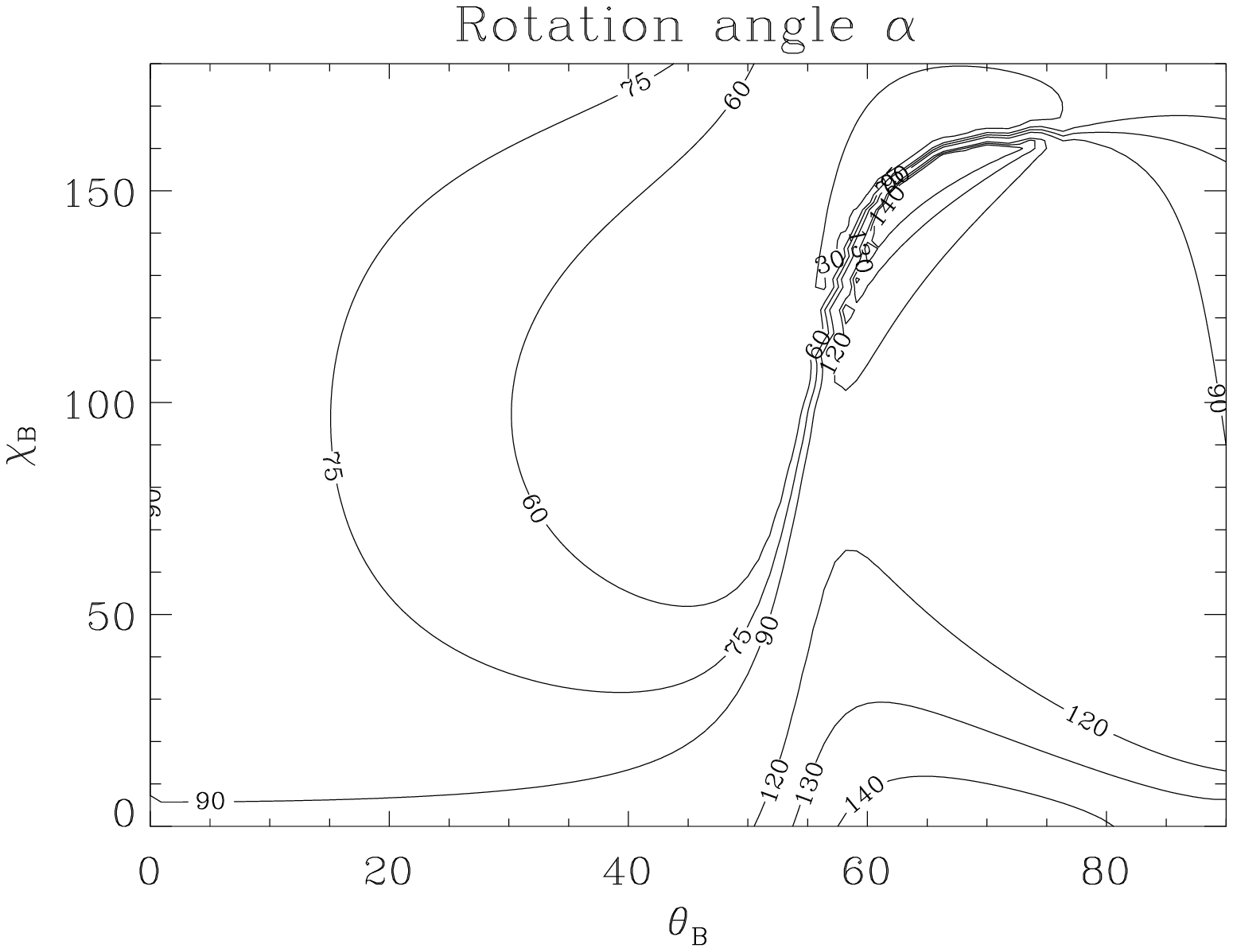}
\caption{Rotation angle defined in Eq. (\ref{eq_rotation_angle}) for different combinations of the field strength and inclination, assuming
$\chi_B=0$ (left panel), and different combinations of the field inclination and azimuth, assuming $\Gamma=10$ (right panel). When the
magnetic field strength is very weak ($\Gamma \to 0$) or it
is along the symmetry axis of the radiation field ($\theta_B=0$) we recover the results obtained for the non-magnetic case, i.e., that
the polarization of the observed light is perpendicular to the quantization axis.
\label{fig_rotation_angle_magnetic}}
\end{figure}

We have shown in the left panel of Figure \ref{fig_rotation_angle_magnetic} the rotation angle when the strength and inclination of
the magnetic field is changed while the azimuth of the field is set to $\chi_B=0$. When the field is weak the rotation angle is close to 90$^\circ$
irrespective of the
value of its inclination, i.e., it is perpendicular to the axis of symmetry of the radiation field.
Moreover, if the field inclination is lower than $\theta_B=54.73^{\circ}$ (the Van Vleck angle), the rotation angle is limited
to 135$^\circ$ irrespective of the value of the field, meaning a rotation of only 45$^\circ$ with respect to the non-magnetic case.
However, when the field increases and the inclination
is above the Van Vleck critical angle, a rotation as large as 90$^\circ$ with respect to the non-magnetic case can be obtained. The
magnetic field strength for obtaining this rotation should be larger than $\sim$90 mG which, in principle, seems quite reasonable.
In the theoretical limit of $\Gamma \to \infty$, there is a sharp division between field inclinations below and above the
Van Vleck angle. When the field inclination is below this critical angle, the polarization angle is 90$^\circ$. When the inclination
is larger than the Van Vleck angle, the polarization angle suddenly changes to 0$^\circ$.

The right panel of Figure \ref{fig_rotation_angle_magnetic} shows the results when $\Gamma=10$ and the inclination and azimuth of the
field are changed. In this case, the rotation angle with respect to the non-magnetic case is larger than 45$^\circ$ only in two very small
regions of the parameter space. One of them is found when the azimuth is close to zero (the same detected in the previous plot) and the other
one is found for azimuths around 150$^\circ$ and inclinations of the order of 65$^\circ$. This last region implies so restrictive conditions
that we consider it not to be responsible for any observable rotation of the direction of polarization.

From the previous analysis, we have seen that there are several combinations of the magnetic field strength and direction (with respect
to the radiation symmetry axis) which gives a rotation of the direction of polarization larger than 45$^\circ$. Although this rotation
could be, in principle, responsible for the rotation of the direction of polarization with respect to the tangential direction
of $\sim 90^\circ$ observed by \citar{kemball_diamond97},
the necessary conditions are quite restrictive.
Moreover, one can note from Figure \ref{fig_maser_condition_magnetic} that these regions are quite coincident with those in which the maser
action is completely inhibited by the presence of the magnetic field.

The investigation of the effect of a magnetic field on the SiO masers allows us to state that, at least under the assumptions we have
made in this modeling, a rotation of more
than 45$^\circ$ of the direction of polarization with respect to the non-magnetic case (i.e. the tangential direction) produced by a
magnetic field is very improbable. In
fact, this
rotation is only obtained under very restricted conditions, which are moreover close to those inducing a complete inhibition of the maser
due to the presence of the
magnetic field. The necessity of this delicate combination of parameters comes from the fact that the magnetic field acts in a twofold
way. On the one hand, it produces a rotation of the polarization angle that can attain very large values ($\sim 90^\circ$) only when
the inclination and strength of the field are increased. On the other hand, the stronger and inclined the field, the larger the
inhibition of the maser effect.


\section{Anisotropy factor}
The previous results suggest that the only effective way of producing a large rotation of the direction of polarization of the maser
radiation with respect to the tangential direction is by a sign change in the anisotropy factor. In this section we calculate the anisotropy factor in a simple model in order to investigate
under which conditions we can have this sign change.

Consider a slab of constant physical properties characterized by the source function $S_0$ which is illuminated by one side by a
collimated and unpolarized radiation field $I_0$. This model represents
the illumination properties of a slab in which the maser is located. The collimated illumination is playing the role of the radiation
field coming from the star which, being quite far away from this region, can be considered to be point-like. Let the total
optical depth of the slab be $2\tau$. In order to calculate the anisotropy factor at the central position of the slab, we have to solve the
radiative transfer equation. Supposing that the illuminating radiation is unpolarized and that the physical conditions are constant
in the slab, we can write the intensity for each angle $\mu=\cos \theta$ as:
\begin{align}
I^+(\tau,\mu) = I_0 \mathrm{e}^{-\tau/\mu} \delta(\mu-1) + S_0 \left( 1- \mathrm{e}^{-\tau/ \mu} \right) \qquad & \mu>0 \nonumber \\
I^-(\tau,\mu) = S_0 \left( 1- \mathrm{e}^{-\tau/ |\mu|} \right) \qquad & \mu<0,
\end{align}
in which the $+$ superscript represents radiation propagating away from the star and the $-$ superscript represents radiation propagating
towards the star. Note that the incoming illuminating radiation contributes only to the radiation with $\mu=1$ because it is
collimated. Once the specific intensity is
obtained we can calculate the tensors of the radiation field given by
Eq. (\ref{eq_tensors_rad_field}) by performing analytically the required angular integrals to end up with the following formula for the anisotropy factor:
\begin{equation}
w = \frac{1}{2} \frac{(\tau^3-2\tau) E_1(\tau) - (\tau^2-\tau-2f) \mathrm{e}^{-\tau} }
{2 + 2\tau E_1(\tau)+(f-2)\mathrm{e}^{-\tau}},
\label{eq_anisot_collimated}
\end{equation}
where $E_1(\tau)$ is the first exponential integral and $f=I_0/S_0$. Note that the anisotropy factor depends only on the total optical
depth of the slab, $\tau$, and on the ratio between the illuminating radiation field and the source
function at the interior of the slab ($I_0/S_0$).

\begin{figure}
\plotone{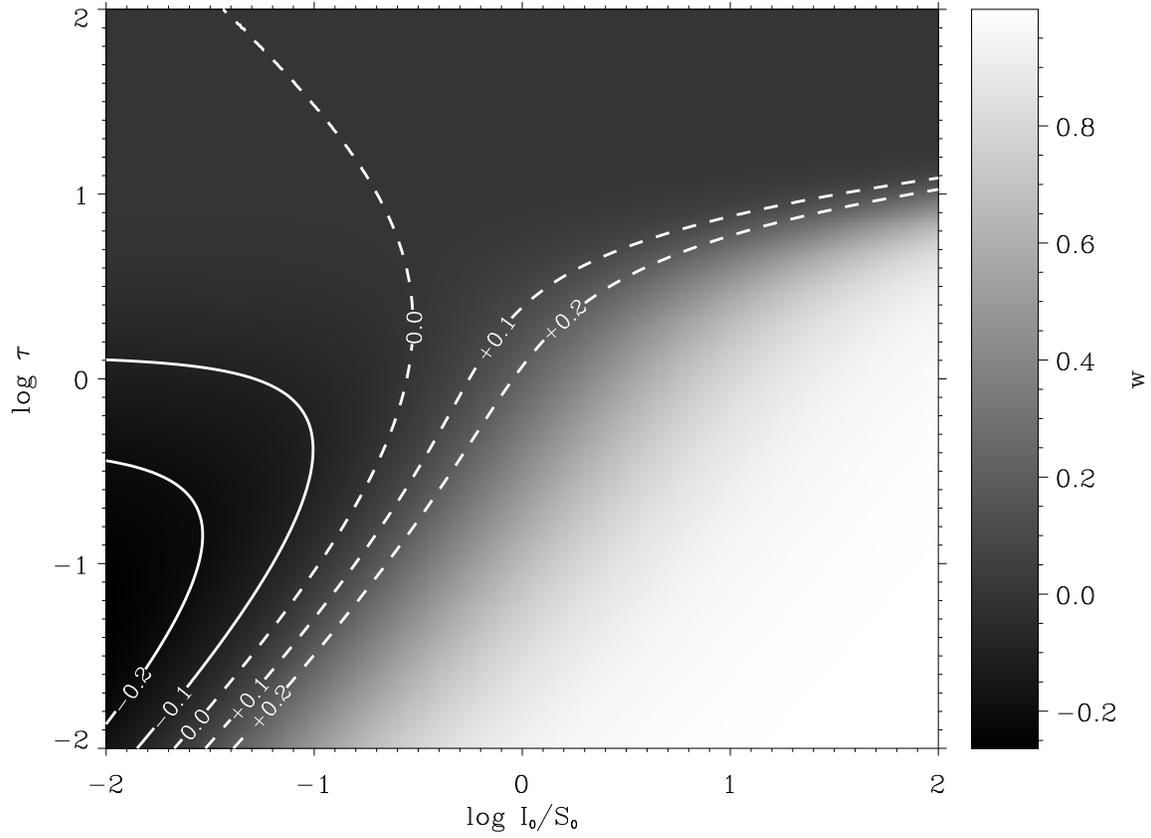}
\caption{Anisotropy factor for different combinations of the total optical depth of the slab and the ratio $I_0/S_0$. For clarity, we
have marked the position of $w=-0.2,-0.1$ with solid lines and the position of $w=0.0,+0.1,+0.2$ with dashed lines.\label{fig_anisotropy}}
\end{figure}

Fig. \ref{fig_anisotropy} shows the value of $w$ for different configurations of the total optical depth of the slab
and the ratio $f$. We have also indicated the curves of constant value of the anisotropy factor. Note that there
is a combination of $\tau$ and $f$ in which the radiation field inside the slab is completely isotropic since $w=0$. For the rest of
combinations, we can find radiation fields inside the slab which are mainly radial ($w>0$) or mainly tangential ($w<0$).
When the source function $S_0$ is more than $\sim 4$ times larger than the incident radiation field $I_0$, both positive and negative
anisotropy factors are possible depending on the optical depth of the slab.

In view of the previous calculations, it is possible to have a change in the sign of the anisotropy factor by local perturbations of
the physical properties of the medium. If the optical depth remains constant and the source function of the slab is increased (for
example due to a local increase in the temperature), there is a transition from a radiation field which is mainly radial (produced
by the illumination of the star) to a radiation field which is mainly tangential (produced by the self-illumination of the
cloud). A similar conclusion was reached by \citar{western_watson83b} by investigating the SiO population inversion obtained
with and without the presence of the stellar radiation field.
On the other hand, if the temperature of the slab remains constant but the optical depth is increased, we can have a similar
transition for a restricted set of ratios between the source function and the incoming radiation field. In case that the
incoming radiation field is very strong, we will always get a radial radiation field (optically thin slab) or an isotropic
field (optically thick slab).

\section{Conclusions}
We have shown that dichroic masing in the unsaturated regime can be considered a very efficient mechanism for producing highly polarized
masers in
diluted media. The nominal value of 100\% for the polarization degree can however be attained only in idealized situations. In
practice, it has to be expected that a number of different phenomena, like trapping of the pumping radiation, the presence of
depolarizing collisions, the effect of saturation (that has been neglected here), and the variation of physical properties along the ray-path
will cooperate in reducing the polarization degree.

Using a simple SiO model we have shown that the rotation of the direction of polarization observed in the maser transitions
in circumstellar envelopes may be produced by the presence of a magnetic field or by a change in the anisotropy of the radiation field.
The rotation due to the presence of an inclined and strong magnetic field is produced under quite restrictive conditions,
since the magnetic field plays two different roles. On the one hand, it brings to a rotation of the direction of polarization. On the other
hand, it produces a strong inhibition effect on the dichroic masers. The conditions under which the rotation of the direction of polarization
of the emerging radiation can be larger than 45$^\circ$ are almost equivalent to the conditions under which a complete inhibition of the
maser effect is produced. Therefore, we consider that the magnetic field is hardly responsible of the observed rotations of the direction
of polarization in circumstellar envelopes. The mechanism which we propose to produce such large rotations is a change
of sign in the anisotropy factor. This change of sign may be produced by a local change in the physical properties of the slab
(e.g., as a result of the presence of shock waves) so that
the radiation illuminating the SiO molecules goes from mainly radial to mainly tangential. We have to remark, however, that in the
saturated regime, a 90$^\circ$ rotation of the plane of polarization can indeed be obtained by a similar rotation of the magnetic
field (\citarNP{goldreich73}) even for isotropic pumping.

\acknowledgments
We thank Valent\'\i n Bujarrabal for stimulating
discussions on the possibility of explaining
the observations of \citar{kemball_diamond97}
via the Hanle effect and to an anonymous referee
for some constructive remarks.
This research has been funded by the
European Commission through the Solar
Magnetism Network (contract HPRN-CT-2002-00313)
and by the
Spanish Ministerio de Educaci\'on y Ciencia
through project AYA2004-05792.

\appendix

\section{Analytical expressions for $f_I$, $f_Q$ and $f_U$}
\label{sec_app_fQ_fU}

These are the quantities appearing in the maser condition when a magnetic field is present. Their analytical expressions are:
\begin{align}
f_I &= (3\mu^2-1) \mathcal{M}^{2}_{00} - 2 \sqrt{6} \mu \sqrt{1-\mu^2} \left( \cos \chi \mathrm{Re} \mathcal{M}^2_{10} - \sin \chi \mathrm{Im} \mathcal{M}^2_{10} \right) +\\
&\sqrt{6} (1-\mu^2) \left( \cos 2\chi \mathrm{Re} \mathcal{M}^2_{20} - \sin 2\chi \mathrm{Im} \mathcal{M}^2_{20} \right) \\
f_Q &= 3(\mu^2-1) \mathcal{M}^2_{00} - 2 \sqrt{6} \mu \sqrt{1-\mu^2} \left( \cos \chi \mathrm{Re} \mathcal{M}^2_{10} - \sin \chi \mathrm{Im} \mathcal{M}^2_{10} \right) -\\
&\sqrt{6} (1+\mu^2) \left( \cos 2\chi \mathrm{Re} \mathcal{M}^2_{20} - \sin 2\chi \mathrm{Im} \mathcal{M}^2_{20} \right) \\
f_U &= 2 \sqrt{6} \sqrt{1-\mu^2} \left( \sin \chi \mathrm{Re} \mathcal{M}^2_{10} + \cos \chi \mathrm{Im} \mathcal{M}^2_{10} \right)
+ 2 \sqrt{6} \mu \left( \sin 2\chi \mathrm{Re} \mathcal{M}^2_{20} + \cos 2\chi \mathrm{Im} \mathcal{M}^2_{20} \right).
\end{align}
In these equations $\mu=\cos \theta$, and the angles $\theta$ (polar angle) and $\chi$ (azimuthal angle) specify the direction of
observation with respect to the $z$-axis.

\clearpage

\clearpage

\clearpage

\clearpage

\clearpage

\clearpage

\clearpage

\clearpage



\begin{thebibliography}{}
\bibitem[Bujarrabal \& Nguyen-Q-Rieu(1981)]{bujarrabal_nguyen81} Bujarrabal, V. \& Nguyen-Q-Rieu 1981, \aap, 102, 65
\bibitem[Bujarrabal(1994)]{bujarrabal94} Bujarrabal, V. 1994, \aap, 285, 953
\bibitem[Casini et al.(2002)]{casini02}
Casini, R., Landi Degl'Innocenti, E., Landolfi, M. \& Trujillo Bueno, J. 2002,
\apj, 573, 864
\bibitem[Cernicharo et al.(1993)]{cerni93} Cernicharo, J., Bujarrabal, V., \& Santar\'en, J. L. 1993, \apj, 407, L33
\bibitem[Davis \& Muenter(1974)]{davis_muenter84} Davis, R. E., \& Muenter, J. S. 1974, J. Chem. Phys., 61, 2940
\bibitem[Desmurs et al.(2000)]{desmurs00} Desmurs, J. F., \& Bujarrabal, V., \& Colomer, F., \& Alcolea, J. 2000, \aap, 360, 189
\bibitem[Drira el al(1997)]{drira97} Drira, I., Hur\'e, J. M., Spielfiedel, A., Feautrier, N., Roueff, E., 1997, \aap, 319, 720
\bibitem[Edmonds(1960)]{edmonds60} Edmonds, A. R., Angular Momentum in Quantum Mechanics (Princeton University Press)
\bibitem[Elitzur(1996)]{elitzur96} Eliztur, M. 1996, \apj, 457, 415
\bibitem[Goldreich, Keeley \& Kwan(1973)]{goldreich73} Goldreich, P., Keeley, D. A., \& Kwan, J. Y. 1973, \apj, 179, 111
\bibitem[Happer(1972)]{happer72} Happer, W. 1972, Rev. Mod. Phys., 44, 169
\bibitem[Jewell et al.(1987)]{jewell87} Jewell, P. R., Dickinson, D. F., Snyder, L. E., Clemens, D. P. 1987, \apj, 323, 749
\bibitem[Kemball \& Diamond(1997)]{kemball_diamond97} Kemball, A. J. \& Diamond, P. J. 1997, \apj, 481, L111
\bibitem[Landi Degl'Innocenti(1982)]{landi82} Landi Degl'Innocenti, E. 1982, \solphys, 79, 291
\bibitem[Landi Degl'Innocenti(1983)]{landi83} Landi Degl'Innocenti, E. 1983, \solphys, 85, 3
\bibitem[Landi Degl'Innocenti(1984)]{landi84} Landi Degl'Innocenti, E. 1984, \solphys, 91, 1
\bibitem[Landi Degl'Innocenti(1998)]{landi98} Landi Degl'Innocenti, E. 1998, Nature, 392, 256
\bibitem[Landi Degl'Innocenti (2003$a$)]{landi03a} Landi Degl'Innocenti, E. 2003$a$,
in {\em Solar Polarization 3}, eds. J. Trujillo Bueno \&
J. S\'anchez Almeida, ASP Conf. Series, Vol. 307, 164
\bibitem[Landi Degl'Innocenti(2003)]{landi03b} Landi Degl'Innocenti, E. 2003$b$,
in {\em Astrophysical Spectropolarimetry}, eds. J. Trujillo Bueno, F. Moreno-Insertis \& F. S\'anchez, Cambridge University Press, 1
\bibitem[Landi Degl'Innocenti \& Landi Degl'Innocenti(1985)]{landi_landi85} Landi Degl'Innocenti, E., \& Landi Degl'Innocenti, M.
1985, \solphys, 95, 239
\bibitem[Landi Degl'Innocenti \& Landolfi(2004)]{landi04} Landi Degl'Innocenti, E., \& Landolfi, M. 2004, Polarization in Spectral
Lines (Kluwer Academic Publishers)
\bibitem[Litvak(1975)]{litvak75} Litvak, M. M. 1975, \apj, 202, 58
\bibitem[Manso Sainz \& Landi Degl'Innocenti(2002)]{manso02}
Manso Sainz, R. \& Landi Degl'Innocenti, E. 2002,
\aap, 394, 1093
\bibitem[Manso Sainz \& Trujillo Bueno(2003)]{manso03}
Manso Sainz, R. \& Trujillo Bueno, J. 2003,
Phys. Rev. Letters, Vol. 91, 111102-1
\bibitem[Trujillo Bueno(1999)]{trujillo99} Trujillo Bueno, J. 1999,
in {\em Solar Polarization}, eds. K.N. Nagendra \& J.O. Stenflo, Kluwer Academic Publishers, 73
\bibitem[Trujillo Bueno(2001)]{trujillo01} Trujillo Bueno, J. 2001,
in {\em Advanced Solar Polarimetry: Theory, Observation and Instrumentation}, ed. M. Sigwarth, ASP Conf. Series, Vol. 236, 161
\bibitem[Trujillo Bueno(2003)]{trujillo03} Trujillo Bueno, J. 2003,
in {\em Stellar Atmosphere Modeling}, eds. I. Hubeny, D. Mihalas \&
K. Werner, ASP Conf. Series, Vol. 288, 551
\bibitem[Trujillo Bueno \& Landi Degl'Innocenti(1997)]{trujillo_landi97} Trujillo Bueno, J., \& Landi Degl'Innocenti, E. 1997,
\apj, 482, L183
\bibitem[Trujillo Bueno et al.(2002$a$)]{trujillo02a} Trujillo Bueno, J.,
Casini, R., Landolfi, M. \& Landi Degl'Innocenti, E. 2002$a$,
\apj, 566, L53
\bibitem[Trujillo Bueno et al.(2002$b$)]{trujillo02b} Trujillo Bueno, J., Landi Degl'Innocenti, E., Collados, M., Merenda, L. \& Manso Sainz, R. 2002$b$,
Nature, 415, 403
\bibitem[Trujillo Bueno et al.(2004)]{trujillo04} Trujillo Bueno, J., Shchukina, N. \& Asensio Ramos, A.  2004,
Nature, 430, 326
\bibitem[Western \& Watson(1983a)]{western_watson83a} Western, L. R. \& Watson, W. D. 1983a, \apj, 268, 849
\bibitem[Western \& Watson(1983b)]{western_watson83b} Western, L. R. \& Watson, W. D. 1983b, \apj, 275, 195
\bibitem[Western \& Watson(1984)]{western_watson84} Western, L. R. \& Watson, W. D. 1984, \apj, 285, 158
\end{thebibliography}
\end{document}